\documentstyle[12pt]{aipproc}

\def\ie{\hbox{\it i.e.}{}}
\def\eg{\hbox{\it e.g.}{}}
\def\nn{\hspace{2mm}}
\def\sss{\scriptscriptstyle}
\def\NPB#1#2#3{Nucl. Phys. {\bf B#1} (#2) #3}
\def\PLB#1#2#3{Phys. Lett. {\bf B#1} (#2) #3}

\def\HPH#1{hep--ph{}/{}#1}

\newcommand{\MeV}{\mbox{\rm MeV}}
\newcommand{\GeV}{\mbox{\rm GeV}}

\def\Tr{\rm{Tr}}

\def\Bar#1{\overline{#1}}

\def\sVEV#1{\left\langle #1\right\rangle}

\def\cL{{\cal L}}
\def\sleq{\raisebox{-.6ex}{${\textstyle\stackrel{<}{\sim}}$}}
\def\sgeq{\raisebox{-.6ex}{${\textstyle\stackrel{>}{\sim}}$}}
\begin{document}
\title{Neutrino Oscillations in Extended Anti-GUT
Model\footnote[3]{Talk given by H.B.~Nielsen at the
Second Tropical Workshop on Particle Physics
and Cosmology, San Juan, Puerto Rico, May 2000.}}
\author{C.D.~Froggatt}
\vspace*{-.5cm}
\address{Department of Physics and Astronomy,\\ Glasgow University,
Glasgow, Scotland}
\author{H.B.~Nielsen}
\vspace*{-.5cm}
\address{Theory Division, CERN\\
and \\
Niels Bohr Institute, Copenhagen {\O}, Denmark}
\author{Y.~Takanishi}
\vspace*{-.5cm}
\address{Niels Bohr Institute, Copenhagen {\O}, Denmark}
\maketitle

\begin{abstract}
What we call the Anti-GUT model is extended a bit to include also
right-handed neutrinos and thus make use of the see-saw mechanism
for neutrino masses. This model consists in assigning gauge quantum numbers
to the known Weyl fermions and the three see-saw right-handed neutrinos.
Each family (generation) is given its own Standard Model gauge fields
and a gauge field coupled to the $B-L$ quantum number
for that family alone. Further we assign a rather
limited number of Higgs fields, so as to break these gauge groups down to the
Standard Model gauge group and to fit, w.r.t. order
of magnitude, the spectra and
mixing angles of the quarks and leptons. We find a rather good fit,
which for neutrino oscillations favours the small mixing angle MSW
solution, although the mixing angle predicted is closest to the upper
side of the uncertainty range for the measured solar neutrino
mixing angle.

An idea for making a ``finetuning''-principle to ``explain'' the large
ratios found empirically in physics, and answer such
questions as ``why is the weak scale low
compared to the Planck scale?'', is proposed.
A further very speculative extension is supposed to
``explain'' why we have three families.
\end{abstract}

\section{Introduction}
Anti-GUT is the name which we have given to the model based on our
favourite gauge group
\begin{equation}
AGUT = SMG\times SMG \times SMG \times U(1)_f.
\end{equation}
It began in the work of one of us (H.B.N.)
with N.~Brene, D.~Bennett and I.~Picek and others, but the
inclusion of the $U(1)_f$ and the application to study the
masses and mixing angles for
quarks and leptons was together with C.D.F.
and his students (G.~Lowe, D.~Smith, M.~Gibson).
Here the symbol $SMG$ stands for the Standard Model Group---and
we may really think about it as a \underline{group}~\cite{or} and
not only a Lie algebra. It may then be assigned physical significance
via the spectrum of representations it has (but one may just think
of it as a Lie algebra, if one wants):
\begin{equation}
SMG = S(U(2)\times U(3)) = SU(2)\times SU(3)\times U(1).
\end{equation}
In the present talk we shall actually present a slightly extended version
of the ``original'' Anti-GUT model,
which could thus be designated the extended
Anti-GUT. This model may be described by saying that to each family
we assign its own set of gauge fields, consisting of a set
of Standard Model gauge fields plus a gauged $U(1)$ group
coupling to the family in question through the $B-L$ quantum number
(but only to that family, the remaining families being considered to have
zero value for this ($B-L$)-charge and for Standard Model ones related to the
considered family gauge bosons). Here
$B-L$ stands for baryon number minus lepton number. That is to say
the gauge group of the extended Anti-GUT---the model we shall consider in the
present talk---is taken to be
\begin{equation}
(SMG\times U(1)_{B-L})^3 %
\approx SU(2)^3\times SU(3)^3\times U(1)^6.
\end{equation}
Here it is meant that the subgroup $(SMG_i\times U(1)_{B-L,i})$
= $S(U(2)\times U(3))\times U(1)$ = $SU(2)\times SU(3) \times U(1)^2$
only couples to the $i$th family and that, in the usual way, the
quarks of family $i$ are triplets under the $SU(3)_i$ (but singlets
under the other $SU(3)_j$'s, $\ie$~when $j\ne i$) and so on. The
$(B-L)_i$ couples to the $B-L$
= baryon number $-$ lepton number for the $i$'th family.

This essentially describes the couplings in the model,
but there is a little tricky point
that was introduced, because it could be shown
that otherwise we would have had a no-go theorem
for fitting the quark spectra~\cite{lowe}:
the right-handed up-type (``up-type''= $u$, $c$, and $t$)
quarks of the second and third family are permuted. That is to say the
right-handed components of what is experimentally seen as the third
family particle---the top quark---are dominantly equal to the right-handed
up-type components coupling to the second family gauge fields.
Similarly the experimentally seen right-handed charm quark is identified
with the formally right-handed top quark of our model, namely the components
coupling to the third family gauge fields.

\section{Motivation for the model}
\label{sec:motivation}
But why should a model of the proposed type be expected to have a chance
of being the right one?

Well, the group may be characterised by means of a few relatively simple
requirements with some phenomenological support: It is the largest
possible gauge group that transforms the known 45 Weyl fermions of the
Standard Model plus three right-handed see-saw neutrinos
into themselves, without unifying any irreducible representations under
the Standard Model and with no anomaly troubles---neither gauge anomalies
nor mixed anomalies. That is to say it may be characterised as the
biggest group for which the gauge symmetry can be upheld
without any Green-Schwarz anomaly cancellation, without unifying particles
that are not already unified in the Standard Model and without
transforming them into ``fantasy particles'' only existing in the model.

Looking for inspiration for what gauge group to
believe in beyond the Standard Model, we are led to consider hints from
the following remarkable features of the quark and lepton spectra:
\begin{list}{$\bullet$}{\usecounter{line}}
\item There are very big mass ratios from family to family and even
to some extent within the families. This feature strongly suggests
that different family particles should have different quantum numbers.
How else should the quantum number differences between right- and
left-handed components be so different that they can be used to
mass protect some particles by much bigger factors than others?

\item Pure $SU(5)$, or for that matter all the Grand Unified Theories (GUTs)
extending $SU(5)$, predicts the
GUT-scale mass relations:
\begin{equation}
\frac{m_{\tau}}{m_{b}}= \frac{m_{\mu}}{m_s} = \frac{m_e}{m_{d}} =1
\label{e4}
\end{equation}
unless one helps it by introducing rather big
representation Higgs fields, \underline{45} say.
However these relations, eq.~(\ref{e4}), are
not well fulfilled by the renormalisation
group evolved experimental values. Only the third family mass ratio
$m_{\tau}/{m_b}$ fits reasonably well with the simple
version of $SU(5)$-GUT (or its SUSY extension).
So we would be much better off w.r.t. fitting
with experiment, if we could get the $SU(5)$ mass predictions only as
predictions up to factors of order of unity,
but not as exact relations.  So we really
do not want to unify the down-type quarks with the charged leptons,
because the masses do not match except in the $\tau-b$ case.
\end{list}

We may classify gauge group proposals, restricted in each case
to the factor group
which acts on the $45$ known Weyl quark and leptons (or $48$
if you count the perhaps non-existing right-handed neutrinos),
according to whether the group is small or big and according to whether
it unifies or avoids unifying the various irreducible representations
under the Standard Model (SM).  Especially we can ask for the four
groups obtained by requiring minimal or maximal degree of such
group size and of the degree of unification. Under all circumstances
we must of course, in order to have a consistent gauge field model,
require that
the anomalies cancel---in the present article we shall ignore the
possibility of Green-Schwarz cancellation, with the excuse that
we do not find any Green-Schwarz anomaly cancellation needed
between the Standard Model gauge fields. Nature has even
``carefully'' chosen to make equally many quark and lepton
families, thereby ensuring simple anomaly cancellation between
the Weyl particle contributions without any
Green-Schwarz mechanism.

The four corners of the set of gauge groups acting on the
Standard Model Weyl fermions can be seen~\cite{Fr}
to correspond to the following gauge groups:
\begin{list}{$\bullet$}{\usecounter{line}}
\item $\underline{\rm small,{}~separating}:$ $SMG$.
\item $\underline{\rm small,{}~unifying}:$   $SU(5)$(= usual GUT).
\item $\underline{\rm large,{}~separating}:$
$SMG^3\times U(1)_f$~(= Anti-GUT)
 \item $\underline{\rm large,{}~unifying}:$
$SU(5)^3\times U(1)_f$~(= Rajpoot Model with $U(1)_f$~\cite{rajpoot}).
\end{list}
\smallskip

If we say that we want many approximately conserved gauge quantum numbers
in order to produce many different big mass
ratios, we are suggested to take ``large'' groups
so as to have many possibilities for separating the particle masses
order of magnitudewise. If we decide that really the mass predictions
in simple GUT-$SU(5)$ are (except for the third family) not
true---because we do not like the
big $\underline{45}$ representation---then we favour a separating
gauge group.
In this way it is suggested that we should seek the right model in the
$\underline{\rm large,{} separating}$-corner; but that is
just our Anti-GUT model!

Originally we started from ideas of what
we called ``confusion''~\cite{nb,book}.
This means that, if there were different gauge fields attached to
isomorphic gauge groups, there is a mechanism (the ``confusion'') that
could cause the gauge group with many
identical cross product factors to break down to its diagonal
subgroup. Thus only one group would survive
from each class of isomorphic ones.
The model used at that time was a chaotic lattice gauge theory
and the breaking to the diagonal subgroup came about by speculating
that,
after some going around in the lattice, one would lose track of which group
was which. In this way there should in practice only be one--the diagonal
subgroup.

Such a model could even ``explain'' why the Standard Model group
is essentially the cross product of the three lowest dimensional
(simple) groups from which to build gauge groups, namely $U(1)$,
$SU(2)$, and $SU(3)$. The explanation should be that on a
very short scale level---Planck scale or fundamental scale say---there
are many gauge groups of a lot of types, but the isomorphic ones get
``confused'' and we only find one representative of each class
of isomorphic ones. Also there is a general expectation that,
in order to survive at scales large compared to the
fundamental scale, the gauge groups should have relatively small
dimension with matter fields in small chiral representations.

We used this picture in connection with the assumption which
we today would call the multiple point principle (MPP~\cite{mpp,mpp1}).
The MPP says that the coupling constants are---by some presumably
yet to be explained effect
(perhaps Baby-universe theory)---being finetuned
by Nature to make a lot of phases meet each other
just for those values of the couplings that are realized in Nature.
This idea of a lot of phases meeting is analogous to the feature of a
microcanonical ensemble that it will often obtain just the temperature
of a phase transition. In English one has a special word for a mixture
of ice and water, namely ``slush''. So such mixtures must be
quite common. When in equilibrium as a microcanonical ensemble,
such slush has (under one atmosphere pressure) always the temperature
that by definition is zero degrees Celsius. The microcanonical ensemble
constitutes a model for how it can happen that this phase
transition temperature $0^{\sss o}$ becomes much more common than
most other single temperatures; there is a delta function
distribution of temperatures with the delta function peaks at the
phase transitions. If there were a true analogy between the
microcanonical ensemble making specific temperatures much more
likely than others and a mechanism making coupling constants in
Nature take those values that are just between the different phases of the
theory, we would have a mechanism for our MPP.

We used a replacement for this slush analogous MPP-idea to
postulate that, in Nature, the gauge couplings should take just
those values that correspond to lattice artifact phase
transitions---values that can be read out of lattice calculation literature.
Combining this postulate (in an old version) with our favourite
Anti-GUT gauge group we fitted the number of families---not
yet known at that time---and tried to say that it was so accurately
determined that we could almost see it was an integer, namely three.
While one of us (H.B.N.) was giving talks about that,
it was found at LEP that
the neutrino decays of the $Z^0$-gauge boson showed that
there were indeed only three families. So we had made a
successful prediction!

\section{Anti-Grand Unification Theory and its extension}
\label{sec:agut}

In this section we review the Anti-GUT model and its
extension describing neutrino masses and mixing angles.

The Anti-GUT model \cite{mark1,mark2,fn,glasgow,fn2,smg3m}
has been put forward by ourselves and collaborators
over many years, with several motivations. It is mainly
justified by a very promising series of experimental
agreements, obtained from fitting
many of the SM parameters with rather few Anti-GUT parameters,
even though most predictions are only made order of
magnitudewise.
\smallskip
The Anti-GUT model deserves its name in as far as its gauge group
$SMG^3\times U(1)_f$, which replaces, so to speak, the often-used GUT
gauge groups such as $SU(5)$, $SO(10)$ {\it etc.}, can
be specified by requiring that:
\begin{enumerate}
\item It should only contain transformations which change the known
45 (= 3 generations of 15 Weyl particles each) Weyl fermions---counted
as left-handed say---into each other unitarily ($\ie$~it must
be a subgroup of $U(45)$).
\item It should be anomaly-free even without using the
Green-Schwarz \cite{gsa} anomaly cancellation mechanism.
\item It should NOT unify the irreducible representations under the
SM gauge group, called here $SMG$ = $SU(3) \times SU(2) \times U(1)$.
\item It should be as big as possible under the foregoing assumptions.
\end{enumerate}

\smallskip

In the present article we shall, however, allow for see-saw
neutrinos---essentially right-handed neutrinos---whereby we want to extend the
number of particles to be transformed under the group being specified
to also include the right-handed neutrinos, even though they have not been
directly ``seen''.

\smallskip

The extended group, which we shall use as
the model gauge group replacing the
unifying groups, can be specified by a similar set of assumptions
to those used above by two of us~\cite{trento} to specify the
``old'' Anti-GUT. We replace assumption $1$ by a
slightly modified assumption which only excludes unobserved fermions
when they have nontrivial quantum numbers under the SM group, so that
they are mass protected. The particles that are mass protected under the
SM would namely be rather light and would likely have been seen. But
see-saw neutrinos with zero SM quantum numbers could not be mass protected by
the SM and could easily be so heavy as not to have been ``seen''.

The model which we have in mind as the extended Anti-GUT model~\cite{NT},
that should
inherit the successes of the ``old'' Anti-GUT model and in addition have
see-saw neutrinos, is proposed to have the gauge group
$SMG^3\times U(1)_f\times U(1)_{\sss\rm B-L,1} \times U(1)_{\sss\rm B-L,23}
\!\approx\!SU(3)^3\times SU(2)^3 \times U(1)^6$. It is assumed to couple
in the following way:

\noindent
The three SM groups $SMG = SU(3) \times SU(2) \times U(1)$
are supposed to be one for each family or generation.
That is to say, for example, there is a first generation $SMG$
among the three; for which all the fermions in the second and third
generations are in the
trivial representations, and with zero charge, while the
first generation particles couple to this first generation $SMG$
as if in the same representations (same charges too) as they are under
the SM. For example, the left proto-electron and the proto-electron
neutrino form a doublet under the $SU(2)_1$ belonging to the first
generation (while they are in singlets w.r.t. the other two $SU(2)$'s)
and have weak hypercharge w.r.t. the first generation $U(1)_1$,
with a value $y_1/2 = -1/2$ analogous to the SM weak hypercharge
being $y/2=-1/2$ for left-handed leptons.

The $U(1)_f$-charge is assigned in a slightly complicated way which is,
however, largely the only one allowed, modulo various permutations and
rewritings, from the no-anomaly requirements. It is zero for all
first-generation particles and for all particles
usually called left-handed. The $U(1)_f$
charge value
on a ``right-handed'' particle
in the second generation is opposite
to that on the corresponding one in
the third generation. See Table $1$ for the detailed assignment.

The two last $U(1)$-groups, $U(1)_{\sss\rm B-L,1}$ and
$U(1)_{\sss\rm B-L,23}$, in our model have charge assignments
corresponding to the quantum number $B-L$ (= baryon number
minus lepton number), though in such a way that the charges
of $U(1)_{\sss\rm B-L,1}$ are zero for the second and third
generations. The $U(1)_{\sss\rm B-L,1}$ charges are
only non-zero for the first generation, for which
they then coincide with the baryon number minus the
lepton number. Analogously the
$U(1)_{\sss\rm B-L,23}$-charge assignments are zero on the
first-generation quarks and leptons, while they coincide
with the baryon number minus the lepton number for second and third
generations. In the next section, we will discuss anomaly
cancellation in the extended Anti-GUT model.

It is then further part of our model that this large gauge group
is broken down spontaneously to the SM group, lying as the
diagonal subgroup of the $SMG^3$ part of the group, by means of a series
of Higgs fields. The quantum numbers of these fields have been
selected mainly from the criterion of fitting the masses and mixing
angles w.r.t. order of magnitude. The Abelian
quantum numbers proposed for
the ``old'' Anti-GUT Higgs fields were:
\noindent
\begin{eqnarray}
&S:& \quad (\frac{1}{6},-\frac{1}{6},0,-1)\\
&W:& \quad (0,-\frac{1}{2},\frac{1}{2},-\frac{4}{3})\\
&T:& \quad (0,-\frac{1}{6},\frac{1}{6}, -\frac{2}{3})\\
&\xi:& \quad (\frac{1}{6},-\frac{1}{6}, 0,0).
\end{eqnarray}%
\noindent
These four Higgs fields are supposed to have VEVs of
the order of between a twentieth and unity compared to the fundamental
scale supposed to be the Planck scale. In addition there was then the
Higgs field under the Anti-GUT-group which should take the role of finally
breaking the SM gauge group down to
$U(3)\!=\! SU(3)\times U(1)_{em}$,~$\ie$ play the role
of the Weinberg-Salam Higgs field:
\begin{equation}
\phi_{WS}: \quad (0, \frac{2}{3},-\frac{1}{6},1).
\end{equation}
Here the quantum numbers were presented in the order of first giving the
three different weak hypercharges corresponding to the three generations
$y_i/2 \,(i=1,2,3)$, and then the $U(1)_f$-charge.

In reference \cite{mark1} we fitted the parameters, being Higgs fields VEVs,
to the masses and mixing angles for charged fermions and the values
are as follows:
\begin{equation}
  \label{eq:vevs}
  \sVEV{S} = 1 \nn,\nn
  \sVEV{W} = 0.179 \nn,\nn
  \sVEV{\xi} = 0.099 \nn,\nn
  \sVEV{T} = 0.071
\end{equation}
In the following we shall often abbreviate the
expressions for these VEVs by deleting
the $\sVEV{\cdots}$ around the Higgs fields,
mostly with the understanding that $S$, $W$, $\dots$
then mean the VEV ``measured in fundamental units''.

In the Anti-GUT model, the old as well as the new, it is assumed
that, at the fundamental (Planck) scale, particles exist
with whatever quantum numbers are needed as propagators in
the chain diagrams used to generate the fermion mass
matrices \cite{fn}, as discussed in section \ref{sec:berechnung}.
The fitted ``suppression factors'' are
the VEVs in units of the ``fundamental scale'' particles.

It has to be checked that extending the group, to have the
$U(1)_{\sss\rm B-L,1}$ and  $U(1)_{\sss\rm B-L,23}$ factors, does not
disturb the successful features of the model.
This can be done by
only giving the fields $\xi$ and $S$ non-zero charges under
these ``new'' $U(1)$ groups, so as to get:
\begin{eqnarray}
&S:&  \quad (\frac{1}{6},-\frac{1}{6},0,-1,-\frac{2}{3},\frac{2}{3})\\
&\xi:& \quad (\frac{1}{6},-\frac{1}{6},0,0,\frac{1}{3},-\frac{1}{3})
\end{eqnarray}%
where the last two quantum numbers are the
$U(1)_{\sss\rm B-L,1}$ and  $U(1)_{\sss\rm B-L,23}$
charges respectively.

But now we also want to introduce two new Higgs fields $\phi_{\sss B-L}$
and $\chi$ into the model: the first, $\phi_{\sss B-L}$, is a Higgs field
used to fit the new scale that comes in
from neutrino oscillations giving the
scale of the see-saw particle masses. When the left-right-transition
(Dirac neutrino)
mass matrix is of the same order as the usual charged fermion mass
matrices, this scale is of the order $10^{12}~\GeV$.

In our model we use the gauged $B-L$, in fact the total one
because we break $U(1)_{\sss B-L,1}\times U(1)_{\sss B-L,23}%
\supseteq U(1)_{{\sss B-L,}{\rm total}}$ at a much higher
scale (near the Planck scale), to mass protect the
right-handed neutrinos.
These right-handed neutrinos
are meant to function as see-saw particles, so
they can be sufficiently light to give
the ``observed'' left-handed neutrino masses
by the see-saw mechanism.  The breaking of the
$U(1)_{{\sss B-L,}{\rm total}}$, and thereby the setting of the
see-saw scale, is then caused by our ``new'' Higgs
field called $\phi_{\sss B-L}$.

In order to get viable neutrino spectra we shall
choose the quantum numbers of $\phi_{\sss B-L}$ so that
the effective
$\Bar{\nu_{\tau_{\sss R}}}\,C\,\Bar{\nu_{e_{\sss R}}}^t + h.c.$ term
gets a direct contribution and is thus not further suppressed.
This is the way to avoid ``factorised mass matrices''---$\ie$~matrices
of the form
\begin{equation}
  \label{eq:factrization}
\left ( \begin{array}{ccc}
        \phi_{1}^2 &\phi_{1}\phi_{2} & \phi_{1}\phi_{3}\\
 \phi_{1}\phi_{2} &\phi_{2}^2 & \phi_{2}\phi_{3}\\
 \phi_{1}\phi_{3} &\phi_{2}\phi_{3} & \phi_{3}^2
                        \end{array} \right )
\end{equation}%
with different order unity factors, though, on different
elements. Such factorised matrices are rather difficult to
avoid otherwise. If we get such a ``factorised matrix'' and, as
in our model, have mainly diagonal elements in the $\nu$-Dirac
matrix, $M_{\nu}^D$, then we get the prediction that
\begin{equation}
\frac{\Delta m^2_{\odot}}{\Delta m^2_{\rm atm}}
\approx (\sin\theta_{\rm atm})^4\nn,
\end{equation}%
which is not true experimentally. Therefore we choose $\phi_{\sss B-L}$
to have the quantum numbers of $\Bar{\nu_{\tau_{\sss R}}}$ plus those of
$\Bar{\nu_{e_{\sss R}}}$:
\begin{eqnarray}
  \label{eq:blladung}
  Q_{\phi_{\sss B-L}} &=& Q_{\bar{\nu}_{\tau_{\sss R}}} + Q_{\bar{\nu}_{e_{\sss
R}}} \nonumber\\
                      &=& (0,0,0,0,1,0) + (0,0,0,1,0,1) \nonumber\\
                      &=& (0,0,0,1,1,1)\nn. \nonumber
\end{eqnarray}
\indent
The other ``new'' Anti-GUT Higgs field we call
$\chi$ and one of its roles is to help
the $\sVEV{\phi_{\sss B-L}}$ to give non-zero effective mass terms for the
see-saw neutrinos, by providing a transition between $\nu_{\tau_{\sss R}}$
and $\nu_{\mu_{\sss R}}$. It also turns out to play a role in fitting
the atmospheric mixing angle (to be of order unity). Its quantum numbers
are therefore postulated to be the difference of those of these two
see-saw particles
\begin{eqnarray}
  \label{eq:chiladung}
  Q_{\chi} &=& Q_{\nu_{\mu_{\sss R}}} - Q_{\nu_{\tau_{\sss R}}} \nonumber\\
           &=& (0,0,0,1,0,-1) - (0,0,0,-1,0,-1) \nonumber\\
           &=& (0,0,0,2,0,0)\nn. \nonumber
\end{eqnarray}
\section{Anomaly Cancellation}
\label{sec:anomaly}
\indent
We introduce here an anomaly-free Abelian
extension of the ``old'' AGUT, which we shall use below
to obtain the neutrino mass spectrum and their mixing angles.
The ``new'' Anti-GUT gauge group is
\begin{equation}
  \label{eq:agutgg}
  SMG^3\times U(1)_f\times U(1)_{\sss B-L,1}\times U(1)_{\sss B-L,23}
\end{equation} and is broken, by a set of Higgs fields
$S$, $W$, $T$, $\xi$, $\chi$ and $\phi_{\sss B-L}$, down to the
SM gauge groups. This diagonal $SMG$ group will finally
be broken down by the field
$\phi_{\sss WS}$, playing the role of the Weinberg-Salam
Higgs field, into $SU(3)\times U(1)_{em}$.

The requirement that all anomalies involving $U(1)_f$,
$U(1)_{\sss\rm B-L,1}$ and $U(1)_{\sss\rm B-L,23}$
then vanish strongly constrains the possible fermion charges (we denote
the $U(1)_f$ charges by $Q_f(t_{\sss R}) \equiv t_R$ {\it etc.} and
the $U(1)_{\sss\rm B-L}$
charges by $Q_{\sss\rm B-L,1}(u_{\sss R}) \equiv \bar{u}_{\sss R}$,
$Q_{\sss\rm B-L,23}(t_{\sss R}) \equiv \tilde{t}_{\sss R}$
{\it etc.} respectively).
The anomaly cancellation conditions constrain the fermion
$U(1)_f$
charges to satisfy the following equations:
\begin{eqnarray}
\Tr\;[SU_1(3)^2 U(1)_f] &=& 2u_{\sss L}-u_{\sss R}-d_{\sss R} =0 \nonumber \\
\Tr\;[SU_2(3)^2 U(1)_f] &=& 2c_{\sss L}-c_{\sss R}-s_{\sss R} =0 \nonumber \\
\Tr\;[SU_3(3)^2 U(1)_f] &=& 2t_{\sss L}-t_{\sss R}-b_{\sss R} =0 \nonumber \\
\Tr\;[SU_1(2)^2 U(1)_f] &=& 3u_{\sss L}+e_{\sss L} =0 \nonumber \\
\Tr\;[SU_2(2)^2 U(1)_f] &=& 3c_{\sss L}+\mu_{\sss L} =0 \nonumber \\
\Tr\;[SU_3(2)^2 U(1)_f] &=& 3t_{\sss L}+\tau_{\sss L} =0 \nonumber \\
\Tr\;[U_1(1)^2 U(1)_f] &=& u_{\sss L} -8u_{\sss R}-2d_{\sss R}+3e_{\sss L}
                         - 6e_{\sss R} =0 \nonumber \\
\Tr\;[U_2(1)^2 U(1)_f] &=& c_{\sss L}-8c_{\sss R}-2s_{\sss R}+3\mu_{\sss L}
                         -6\mu_{\sss R} =0\nonumber \\
\Tr\;[U_3(1)^2 U(1)_f] &=& t_{\sss L}-8t_{\sss R}-2b_{\sss R}+3\tau_{\sss L}
                         -6\tau_{\sss R} =0\nonumber \\
\Tr\;[U_1(1) U(1)_f^2] &=& u_{\sss L}^2-2u_{\sss R}^2+d_{\sss R}^2-e_{\sss L}^2
                         +e_{\sss R}^2 =0\nonumber \\
\Tr\;[U_2(1) U(1)_f^2] &=& c_{\sss L}^2-2c_{\sss R}^2
+s_{\sss R}^2-\mu_{\sss L}^2
                         +\mu_{\sss R}^2 =0\nonumber \\
\Tr\;[U_3(1) U(1)_f^2] &=& t_{\sss L}^2-2t_{\sss R}^2
+b_{\sss R}^2-\tau_{\sss L}^2
                         +\tau_{\sss R}^2=0 \nonumber \\
\Tr\;[U(1)_f^3] &=& 6u_{\sss L}^3+6c_{\sss L}^3
+6t_{\sss L}^3-3u_{\sss R}^3-3c_{\sss R}^3
                  -3t_{\sss R}^3-3d_{\sss R}^3-3s_{\sss R}^3 \nonumber \\&&
             -3b_{\sss R}^3+2e_{\sss L}^3+2\mu_{\sss L}^3 +2\tau_{\sss L}^3
                   -e_{\sss R}^3-\mu_{\sss R}^3-\tau_{\sss R}^3 \nonumber\\
           && -\nu_{e_{\sss R}}^3-\nu_{\mu_{\sss R}}^3
-\nu_{\tau_{\sss R}}^3 = 0 \nonumber \\
\Tr\;[{(\rm graviton)}^2 U(1)_f] &=& 6u_{\sss L}+6c_{\sss L}
+6t_{\sss L}-3u_{\sss R}
-3c_{\sss R}-3t_{\sss R}-3d_{\sss R}-3s_{\sss R} \nonumber \\
&& -3b_{\sss R}+2e_{\sss L}+2\mu_{\sss L}
+2\tau_{\sss L}-e_{\sss R}-\mu_{\sss R}
-\tau_{\sss R}\nonumber\\
&& -\nu_{e_{\sss R}}-\nu_{\mu_{\sss R}}-\nu_{\tau_{\sss R}}=0 \nonumber
\end{eqnarray}%
Similar conditions should be obeyed replacing $U(1)_f$ both by
$U(1)_{\sss B-L,1}$, and by $U(1)_{\sss B-L,23}$,
$\ie$~replacing the $t_{\sss R}$, $b_{\sss R}$, $\dots$ by
$\tilde{t}_{\sss R}$, $\tilde{b}_{\sss R}$, $\dots${}:
\begin{eqnarray}
\Tr\;[SU_1(3)^2 U(1)_{\rm\sss B-L,1}] &=& 2\bar{u}_{\sss L}
-\bar{u}_{\sss R}-\bar{d}_{\sss R} =0 \nonumber \\
\Tr\;[SU_2(3)^2 U(1)_{\rm\sss B-L,23}] &=& 2\tilde{c}_{\sss L}
-\tilde{c}_{\sss R}-\tilde{s}_{\sss R} =0 \nonumber \\
\Tr\;[SU_3(3)^2 U(1)_{\rm\sss B-L,23}] &=& 2\tilde{t}_{\sss L}
-\tilde{t}_{\sss R}-\tilde{b}_{\sss R} =0 \nonumber \\
\Tr\;[SU_1(2)^2 U(1)_{\rm\sss B-L,1}] &=& 3\bar{u}_{\sss L}
+\bar{e}_{\sss L} =0 \nonumber \\
\Tr\;[SU_2(2)^2 U(1)_{\rm\sss B-L,23}] &=& 3\tilde{c}_{\sss L}
+\tilde{\mu}_{\sss L} =0 \nonumber \\
\Tr\;[SU_3(2)^2 U(1)_{\rm\sss B-L,23}] &=& 3\tilde{t}_{\sss L}
+\tilde{\tau}_{\sss L} =0 \nonumber
\end{eqnarray}
But with several $U(1)$s, there will in addition be anomaly conditions
for combinations between the different ones. Taking into account that
$U(1)_{\sss B-L,1}$ charges are zero for all second- and third-generation
fermions, while $U(1)_{\sss B-L,23}$ charges are
zero for the first generation, the further conditions are:
\begin{eqnarray}
\Tr\;[U_1(1)^2 U(1)_{\rm\sss B-L,1}] &=& \bar{u}_{\sss L}
-8\bar{u}_{\sss R}-2\bar{d}_{\sss R}+3\bar{e}_{\sss L}
                         - 6\bar{e}_{\sss R} =0 \nonumber \\
\Tr\;[U_2(1)^2 U(1)_{\rm\sss B-L,23}] &=&
\tilde{c}_{\sss L}-8\tilde{c}_{\sss R}-2\tilde{s}_{\sss R}
+3\tilde{\mu}_{\sss L} -6\tilde{\mu}_{\sss R} =0\nonumber \\
\Tr\;[U_3(1)^2 U(1)_{\rm\sss B-L,23}] &=& \tilde{t}_{\sss L}
-8\tilde{t}_{\sss R}-2\tilde{b}_{\sss R}
+3\tilde{\tau}_{\sss L}-6\tilde{\tau}_{\sss R} =0\nonumber \\
\Tr\;[U_1(1) U(1)_{\rm\sss B-L,1}^2] &=& \bar{u}_{\sss L}^2
-2\bar{u}_{\sss R}^2+\bar{d}_{\sss R}^2-\bar{e}_{\sss L}^2
                         +\bar{e}_{\sss R}^2 =0\nonumber \\
\Tr\;[U_2(1) U(1)_{\rm\sss B-L,23}^2] &=& \tilde{c}_{\sss L}^2
-2\tilde{c}_{\sss R}^2+\tilde{s}_{\sss R}^2-\tilde{\mu}_{\sss L}^2
+\tilde{\mu}_{\sss R}^2 =0\nonumber \\
\Tr\;[U_3(1) U(1)_{\rm\sss B-L,23}^2] &=& \tilde{t}_{\sss L}^2
-2\tilde{t}_{\sss R}^2+\tilde{b}_{\sss R}^2-\tilde{\tau}_{\sss L}^2
+\tilde{\tau}_{\sss R}^2=0 \nonumber \\
\Tr\;[U(1)_{\rm\sss B-L,1}^3] &=& 6\bar{u}_{\sss L}^3
-3\bar{u}_{\sss R}^3-3\bar{d}_{\sss R}^3+2\bar{e}_{\sss L}^3
-\bar{e}_{\sss R}^3-\bar{\nu}_{e_{\sss R}}^3=0 \nonumber \\
\Tr\;[U(1)_{\rm\sss B-L,23}^3] &=& 6\tilde{c}_{\sss L}^3
+6\tilde{t}_{\sss L}^3-3\tilde{c}_{\sss R}^3-3\tilde{t}_{\sss R}^3
-3\tilde{s}_{\sss R}^3-3\tilde{b}_{\sss R}^3\nonumber\\
&&  +2\tilde{\mu}_{\sss L}^3 +2\tilde{\tau}_{\sss L}^3
-\tilde{\mu}_{\sss R}^3-\tilde{\tau}_{\sss R}^3
-\tilde{\nu}_{\mu_{\sss R}}^3-\tilde{\nu}_{\tau_{\sss R}}^3=0 \nonumber \\
\Tr\;[{(\rm graviton)}^2 U(1)_{\rm\sss B-L,1}] &=& 6\bar{u}_{\sss L}
-3\bar{u}_{\sss R}-3\bar{d}_{\sss R}+2\bar{e}_{\sss L}-\bar{e}_{\sss R}
-\bar{\nu}_{e_{\sss R}}=0 \nonumber\\
\Tr\;[{(\rm graviton)}^2 U(1)_{\rm\sss B-L,23}] &=& 6\tilde{c}_{\sss L}
+6\tilde{t}_{\sss L}-3\tilde{c}_{\sss R}-3\tilde{t}_{\sss R}
-3\tilde{s}_{\sss R}-3\tilde{b}_{\sss R}\nonumber\\
&&  +2\tilde{\mu}_{\sss L}+2\tilde{\tau}_{\sss L}-\tilde{\mu}_{\sss R}
-\tilde{\tau}_{\sss R}-\tilde{\nu}_{\mu_{\sss R}}
-\tilde{\nu}_{\tau_{\sss R}}=0 \nonumber\\
\Tr\;[U(1)_f^2 U(1)_{\sss B-L,1}] &=&
6u_{\sss L}^2\bar{u}_{\sss L}-3u_{\sss R}^2\bar{u}_{\sss R}
-3d_{\sss R}^2\bar{d}_{\sss R} %
+2e_{\sss L}^2\bar{e}_{\sss L}- e_{\sss R}^2\bar{e}_{\sss R}
-\nu_{e_{\sss R}}^2\bar{\nu}_{e_{\sss R}} = 0\nonumber\\
\Tr\;[U(1)_f^2 U(1)_{\sss B-L,23}] &=&  6c_{\sss L}^2\tilde{c}_{\sss L}
-3c_{\sss R}^2\tilde{c}_{\sss R}-3s_{\sss R}^2\tilde{s}_{\sss R}%
+ 2\mu_{\sss L}^2\tilde{\mu}_{\sss L}- \mu_{\sss R}^2\tilde{\mu}_{\sss R}
-\nu_{\mu_{\sss R}}^2\tilde{\nu}_{\mu_{\sss R}}\nonumber\\
&& +6t_{\sss L}^2\tilde{t}_{\sss L}-3t_{\sss R}^2\tilde{t}_{\sss R}
-3b_{\sss R}^2\tilde{b}_{\sss R} %
+ 2\tau_{\sss L}^2\tilde{\tau}_{\sss L} - \tau_{\sss R}^2\tilde{\tau}_{\sss R}
-\nu_{\tau_{\sss R}}^2\tilde{\nu}_{\tau_{\sss R}} = 0\nonumber\\
\Tr\;[U(1)_f U(1)_{\sss B-L,1}^2] &=&
6u_{\sss L}\bar{u}_{\sss L}^2-3u_{\sss R}\bar{u}_{\sss R}^2
-3d_{\sss R}\bar{d}_{\sss R}^2 %
+2e_{\sss L}\bar{e}_{\sss L}^2- e_{\sss R}\bar{e}_{\sss R}^2
-\nu_{e_{\sss R}}\bar{\nu}_{e_{\sss R}}^2 = 0\nonumber\\
\Tr\;[U(1)_f U(1)_{\sss B-L,23}^2] &=&  6c_{\sss L}\tilde{c}_{\sss L}^2
-3c_{\sss R}\tilde{c}_{\sss R}^2-3s_{\sss R}\tilde{s}_{\sss R}^2%
+ 2\mu_{\sss L}\tilde{\mu}_{\sss L}^2- \mu_{\sss R}\tilde{\mu}_{\sss R}^2
-\nu_{\mu_{\sss R}}\tilde{\nu}_{\mu_{\sss R}}^2\nonumber\\
&& +6t_{\sss L}\tilde{t}_{\sss L}^2-3t_{\sss R}\tilde{t}_{\sss R}^2
-3b_{\sss R}\tilde{b}_{\sss R}^2 %
+ 2\tau_{\sss L}\tilde{\tau}_{\sss L}^2 - \tau_{\sss R}\tilde{\tau}_{\sss R}^2
-\nu_{\tau_{\sss R}}\tilde{\nu}_{\tau_{\sss R}}^2 = 0\nonumber\\
\Tr\;[U(1)_1 U(1)_f U(1)_{\sss B-L,1}] &=& u_{\sss L}\bar{u}_{\sss L}
-2 u_{\sss R}\bar{u}_{\sss R} + d_{\sss R}\bar{d}_{\sss R}
- e_{\sss L}\bar{e}_{\sss L}+ e_{\sss R}\bar{e}_{\sss R} = 0\nonumber\\
\Tr\;[U(1)_2 U(1)_f U(1)_{\sss B-L,23}] &=&  c_{\sss L}\tilde{c}_{\sss L}
-2 c_{\sss R}\tilde{c}_{\sss R} + s_{\sss R}\tilde{s}_{\sss R}
- \mu_{\sss L}\tilde{\mu}_{\sss L}
+ \mu_{\sss R}\tilde{\mu}_{\sss R} = 0\nonumber\\
\Tr\;[U(1)_3 U(1)_f U(1)_{\sss B-L,23}] &=&  t_{\sss L}\tilde{t}_{\sss L}
-2 t_{\sss R}\tilde{t}_{\sss R} + b_{\sss R}\tilde{b}_{\sss R}
- \tau_{\sss L}\tilde{\tau}_{\sss L}
+ \tau_{\sss R}\tilde{\tau}_{\sss R} = 0\nonumber
\end{eqnarray}

\vspace*{.5cm}

\begin{table}[!tt]
\caption{All $U(1)$ quantum charges in the extended Anti-GUT model.}
\label{Table1}
\begin{center}
\begin{tabular}{|c||c|c|c|c|c|c|} \hline
& $SMG_1$& $SMG_2$ & $SMG_3$ & $U(1)_f$ & $U_{\sss B-L,1}$ &
$U_{\sss B-L,23}$ \\ \hline\hline
$u_L,d_L$ &  $\frac{1}{6}$ & $0$ & $0$ & $0$ & $\frac{1}{3}$ & $0$ \\
$u_R$ &  $\frac{2}{3}$ & $0$ & $0$ & $0$ & $\frac{1}{3}$ & $0$ \\
$d_R$ & $-\frac{1}{3}$ & $0$ & $0$ & $0$ & $\frac{1}{3}$ & $0$ \\
$e_L, \nu_{e_{\sss L}}$ & $-\frac{1}{2}$ & $0$ & $0$ & $0$ & $-1$ & $0$ \\
$e_R$ & $-1$ & $0$ & $0$ & $0$ & $-1$ & $0$ \\
$\nu_{e_{\sss R}}$ &  $0$ & $0$ & $0$ & $0$ & $-1$ & $0$ \\ \hline
$c_L,s_L$ & $0$ & $\frac{1}{6}$ & $0$ & $0$ & $0$ & $\frac{1}{3}$ \\
$c_R$ &  $0$ & $\frac{2}{3}$ & $0$ & $1$ & $0$ & $\frac{1}{3}$ \\
$s_R$ & $0$ & $-\frac{1}{3}$ & $0$ & $-1$ & $0$ & $\frac{1}{3}$\\
$\mu_L, \nu_{\mu_{\sss L}}$ & $0$ & $-\frac{1}{2}$ & $0$ & $0$ & $0$ & $-1$\\
$\mu_R$ & $0$ & $-1$ & $0$ & $-1$  & $0$ & $-1$ \\
$\nu_{\mu_{\sss R}}$ &  $0$ & $0$ & $0$ & $1$ & $0$ & $-1$ \\ \hline
$t_L,b_L$ & $0$ & $0$ & $\frac{1}{6}$ & $0$ & $0$ & $\frac{1}{3}$ \\
$t_R$ &  $0$ & $0$ & $\frac{2}{3}$ & $-1$ & $0$ & $\frac{1}{3}$ \\
$b_R$ & $0$ & $0$ & $-\frac{1}{3}$ & $1$ & $0$ & $\frac{1}{3}$\\
$\tau_L, \nu_{\tau_{\sss L}}$ & $0$ & $0$ & $-\frac{1}{2}$ & $0$ & $0$ & $-1$\\
$\tau_R$ & $0$ & $0$ & $-1$ & $1$ & $0$ & $-1$\\
$\nu_{\tau_{\sss R}}$ &  $0$ & $0$ & $0$ & $-1$ & $0$ & $-1$ \\ \hline \hline
$\phi_{\sss WS}$ & $0$ & $\frac{2}{3}$ & $-\frac{1}{6}$ & $1$ & $0$ & $0$ \\
$S$ & $\frac{1}{6}$ & $-\frac{1}{6}$ & $0$ & $-1$ & $-\frac{2}{3}$ &
$\frac{2}{3}$\\
$W$ & $0$ & $-\frac{1}{2}$ & $\frac{1}{2}$ & $-\frac{4}{3}$ & $0$ & $0$ \\
$\xi$ & $\frac{1}{6}$ & $-\frac{1}{6}$ & $0$ & $0$ & $\frac{1}{3}$ & $
-\frac{1}{3}$\\
$T$ & $0$ & $-\frac{1}{6}$ & $\frac{1}{6}$ & $-\frac{2}{3}$ & $0$ & $0$\\
$\chi$ & $0$ & $0$ & $0$ & 2 & $0$ & $0$ \\
$\phi_{\sss B-L}$ & $0$ & $0$ & $0$ & $1$ & $1$ & $1$ \\ \hline
\end{tabular}
\end{center}
\end{table}

{}From these equations we can get the following solutions:
\begin{eqnarray}
(u_{\sss L},u_{\sss R},d_{\sss R},e_{\sss L},e_{\sss R},\nu_{e_{\sss R}})
&=& (0,0,0,0,0,0) \nonumber \\
(c_{\sss L},c_{\sss R},s_{\sss R},\mu_{\sss L},\mu_{\sss R},
\nu_{\mu_{\sss R}}) &=& (0,1,-1,0,-1,1) \nonumber \\
(t_{\sss L},t_{\sss R},b_{\sss R},\tau_{\sss L},\tau_{\sss R},
\nu_{\tau_{\sss R}}) &=& (0,-1,1,0,1,-1)\nonumber\\
(\bar{u}_{\sss L}, \bar{u}_{\sss R}, \bar{d}_{\sss L}, \bar{d}_{\sss R},
\bar{e}_{\sss L}, \bar{e}_{\sss R}, \bar{\nu}_{e_{\sss L}},
\bar{\nu}_{e_{\sss R}}) &=& (\frac{1}{3},\frac{1}{3},\frac{1}{3},
\frac{1}{3}, -1,-1,-1,-1) \nonumber\\
(\tilde{c}_{\sss L}, \tilde{c}_{\sss R}, \tilde{s}_{\sss L},
\tilde{s}_{\sss R},\tilde{b}_{\sss L}, \tilde{b}_{\sss R},
\tilde{t}_{\sss L}, \tilde{t}_{\sss R}) &=& (\frac{1}{3},
\frac{1}{3},\frac{1}{3},\frac{1}{3},\frac{1}{3},
\frac{1}{3},\frac{1}{3},\frac{1}{3}) \nonumber\\
(\tilde{\mu}_{\sss L}, \tilde{\mu}_{\sss R}, \tilde{\tau}_{\sss L},
\tilde{\tau}_{\sss R},\tilde{\nu}_{\mu_{\sss L}},
\tilde{\nu}_{\mu_{\sss R}}, \tilde{\nu}_{\tau_{\sss L}},
\tilde{\nu}_{\tau_{\sss R}}) &=& (-1,-1,-1,-1,-1,-1,-1,-1)\nonumber
\end{eqnarray}

We summarise the Abelian gauge quantum numbers of our model for
fermions and scalars in Table $1$. However, the following
three points should be kept in mind; then the information
in Table $1$ and these three points describe our whole model:
\begin{enumerate}
\item We have only presented here the six $U(1)$-charges in our
model. The non-Abelian quantum charge numbers are to be derived
from the following rule:
\smallskip%
find in the table $y_i/2$ ($i=1,2,3$ is the generation number), then
find that Weyl particle in the SM for which the SM weak hypercharge
divided by two is $y/2=y_i/2$ and use its $SU(2)$ and $SU(3)$
representation for the particle considered in the
table. But now use it for $SU(2)_i$ and $SU(3)_i$.
\item Remember we imagine that at the ``fundamental'' scale
(presumed to be $\simeq$ the Planck scale)
we have essentially all particles that can
be imagined with couplings of order unity. But we do not want to be specific
about these very heavy particles, in order not to decrease the
likelihood of our model being right. We are only specific
about the particles in Table \ref{Table1} and the gauge fields.
\item The $39$ gauge bosons correspond to the group (equation
(\ref{eq:agutgg})) and are not included in Table \ref{Table1}.
\end{enumerate}


\section{Improved charge formulation}

The system of charges just presented may seem a little complicated
and arbitrary. However the fermion charge combinations are so restricted,
by the anomaly conditions and the connection to the Standard Model,
that they can essentially only be permuted in various ways. We can
though transform these charges into some linear combinations that
come to look nicer and easier to remember; but the physical content
of the theory is of course the same in the reformulated version.

Indeed it turns out that the $U(1)_f$-charge, $Q_f$,
contains the information which corresponds
to letting even the second and third families have their
separate ($B-L$)-charges. We can define generally the
second and third family ($B-L$)-charges:
\begin{equation}
(B-L)_2 = \frac{1}{2} (B-L)_{23} +\frac{y_2}{2} - \frac{y_3}{2} -
\frac{Q_f}{2}
\label{BL2}
\end{equation}
and
\begin{equation}
(B-L)_3 = \frac{1}{2} (B-L)_{23} +\frac{y_3}{2} - \frac{y_2}{2} +
\frac{Q_f}{2}
\label{BL3}
\end{equation}
for the Weyl particles (the fermions in the Standard Model
with the charges in our scheme). The $(B-L)_i$ charges then
take on their well-known values as given in Table~\ref{Table2}.
So they should be no problem to remember and one can easily reconstruct the
$U(1)_f$-charges, from the above two formulae
(\ref{BL2},\ref{BL3}), in case one should want them.

One can also formally use the same formulae for the
quantum numbers of the Higgs fields which we have proposed
and obtain their values in the new notation, as given in
Table~\ref{Table2}.

\begin{table}[!tt]
\caption{All $U(1)$ quantum charges in the re-extended Anti-GUT model.}
\label{Table2}
\begin{center}
\begin{tabular}{|c||c|c|c|c|c|c|} \hline
& $SMG_1$& $SMG_2$ & $SMG_3$ & $U_{\sss B-L,1}$ & $U_{\sss B-L,2}$ &
$U_{\sss B-L,3}$ \\ \hline\hline
$u_L,d_L$ &  $\frac{1}{6}$ & $0$ & $0$ & $\frac{1}{3}$ & $0$ & $0$ \\
$u_R$ &  $\frac{2}{3}$ & $0$ & $0$ & $\frac{1}{3}$ & $0$ & $0$ \\
$d_R$ & $-\frac{1}{3}$ & $0$ & $0$ & $\frac{1}{3}$ & $0$ & $0$ \\
$e_L, \nu_{e_{\sss L}}$ & $-\frac{1}{2}$ & $0$ & $0$ & $-1$ & $0$ & $0$ \\
$e_R$ & $-1$ & $0$ & $0$ & $-1$ & $0$ & $0$ \\
$\nu_{e_{\sss R}}$ &  $0$ & $0$ & $0$ & $-1$ & $0$ & $0$ \\ \hline
$c_L,s_L$ & $0$ & $\frac{1}{6}$ & $0$ & $0$ & $\frac{1}{3}$ & $0$ \\
$c_R$ &  $0$ & $\frac{2}{3}$ & $0$ & $0$ & $\frac{1}{3}$ & $0$ \\
$s_R$ & $0$ & $-\frac{1}{3}$ & $0$ & $0$ & $\frac{1}{3}$ & $0$\\
$\mu_L, \nu_{\mu_{\sss L}}$ & $0$ & $-\frac{1}{2}$ & $0$ & $0$ & $-1$ &
$0$\\
$\mu_R$ & $0$ & $-1$ & $0$ & $0$  & $-1$ & $0$ \\
$\nu_{\mu_{\sss R}}$ &  $0$ & $0$ & $0$ & $0$ & $-1$ & $0$ \\ \hline
$t_L,b_L$ & $0$ & $0$ & $\frac{1}{6}$ & $0$ & $0$ & $\frac{1}{3}$ \\
$t_R$ &  $0$ & $0$ & $\frac{2}{3}$ & $0$ & $0$ & $\frac{1}{3}$ \\
$b_R$ & $0$ & $0$ & $-\frac{1}{3}$ & $0$ & $0$ & $\frac{1}{3}$\\
$\tau_L, \nu_{\tau_{\sss L}}$ & $0$ & $0$ & $-\frac{1}{2}$ & $0$ & $0$ &
$-1$\\
$\tau_R$ & $0$ & $0$ & $-1$ & $0$ & $0$ & $-1$\\
$\nu_{\tau_{\sss R}}$ &  $0$ & $0$ & $0$ & $0$ & $0$ & $-1$ \\ \hline
\hline
$\phi_{\sss WS}$ & $0$ & $\frac{2}{3}$ & $-\frac{1}{6}$ & $0$ &
$\frac{1}{3}$ & $-\frac{1}{3}$ \\
$S$ & $\frac{1}{6}$ & $-\frac{1}{6}$ & $0$ & $-\frac{2}{3}$ &
$\frac{2}{3}$ & $0$\\
$W$ & $0$ & $-\frac{1}{2}$ & $\frac{1}{2}$ & $0$ & $-\frac{1}{3}$ &
$\frac{1}{3}$ \\
$\xi$ & $\frac{1}{6}$ & $-\frac{1}{6}$ & $0$ & $\frac{1}{3}$ &
$-\frac{1}{3}$ & $0$\\
$T$ & $0$ & $-\frac{1}{6}$ & $\frac{1}{6}$ & $0$ & $0$ & $0$\\
$\chi$ & $0$ & $0$ & $0$ & $0$ & $-1$ & $1$ \\
$\phi_{\sss B-L}$ & $0$ & $0$ & $0$ & $1$ & $0$ & $1$ \\ \hline
\end{tabular}
\end{center}
\end{table}

A technical detail that can be useful in checking our tables and
the mass matrices is the regularity
\begin{equation}
(B-L)_i = -\frac{y_i}{2} \quad (\rm{mod}\; \frac{1}{2})
\quad {\rm for} \quad $i$=1,2,3.
\label{modhalf}
\end{equation}
which follows from the facts that the quarks have $(B-L)_i =1/3$
for their family $i$ say (and zero for the other family $(B-L)_j$ 's)
and that their family weak hypercharge satisfies
\begin{equation}
\frac{y_i}{2} = - \frac{\text{``triality''}}{3}\quad
(\rm{mod}\; \frac{1}{2})
\end{equation}
As one can see from Table~\ref{Table2},
most of the Higgs fields which we have
proposed also have quantum number assignments obeying the rules of
eq.~(\ref{modhalf}), but the two Higgs
fields $W$ and $T$ do \underline{not}
obey the rules. In fact they do obey the rule for the first family
$\ie$~for $i=1$, but have  deviations of just opposite sign
for the second and third family---so that the
total $B-L$ and the total weak hypercharge
$\frac{y}{2}$ obey the rule (\ref{modhalf})
even for the $W$ and $T$ fields.
In fact $W$ has $(B-L)_2 + y_2/2 = -1/3 + (-1/2)
= 1/6\,(\rm{mod}\, 1/2)$ and $(B-L)_3 + y_3/2 = \frac{1}{3} + 1/2
= 1/3\, (\rm{mod}\, 1/2)$, while $T$ has
$(B-L)_2 + y_2/2 =0 + (-1/6) = 1/3 (\rm{mod}\, 1/2)$
and $(B-L)_3 + y_3/2 = 0 + 1/6 = 1/6 (\rm{mod}\, 1/2)$.
It then follows with the above assignment of charges that, in
order to make transitions between left- and right-handed quarks or
leptons,
the fields $W$ and $T$ can only occur
in such combinations that the quantities $(B-L)_i + y_i/2$ add
up to zero modulo $1/2$. That is to say combinations like
$T^3$, $T^2W^\dagger$, $T(W^\dagger)^2$, $(W^\dagger)^3$,
$T^\dagger W^\dagger$, ... are allowed as well as their
conjugates, but ~$\eg$~just $T$ or just $W$ or $W^\dagger$
would not be allowed to occur as mass matrix elements.

\section{Mass matrices within the Anti-GUT model}
\label{sec:massmatrizen}

In the ``old'' Anti-GUT model we have only the usual SM fermions at
low energies, but in our ``new'' version we assume that
there exist very heavy right-handed neutrinos,
all of them having already decayed and not being observable in
our world. They function as see-saw particles
and thus give rise to an effective Majorana-type mass matrix for the
left-handed neutrinos. These three ``right-handed''neutrinos
get masses from the VEV of $\phi_{\sss B-L}$ $(10^{12}~\GeV)$, $\xi$ and
also $\chi$ Higgs fields (the latter two
having a VEV of order $1/10$ in Planck units).

The effective mass matrix elements, left-left, for the left-handed
neutrinos---the ones we ``see'' experimentally---then come about using
the $\nu_R$ see-saw propagator surrounded by left-right transition
neutrino mass matrices. The latter are rather analogous to
the charged lepton and quark mass matrices which are proportional
to the VEV of the Weinberg-Salam Higgs field,
being components of $\phi_{\sss WS}$ (with VEV $\sim 173~\GeV$)
in our model.

\smallskip

The Higgs field $\phi_{\sss\rm B-L}$ breaks the total $B-L$
quantum number, as well as the first and third family
$(B-L)_i$ quantum numbers.
Thus the effective Majorana mass terms
are added into the Lagrange density using the Higgs field
$\phi_{\sss B-L}$. The part of the effective Lagrangian we
have to consider is:
\begin{eqnarray}
  \label{lagrangian}
-\cL_{\sss\rm lepton-mass} &\!=\!&
\bar{\nu}_L \,M^D_\nu \,\nu_R
+  \frac{1}{2}(\Bar{\nu_L})^{\sss c}\,M_{L}\, \nu_L
+  \frac{1}{2}(\Bar{\nu_R})^{\sss c}\,M_{R}\, \nu_R + h.c. \nonumber\\
&\!=\!&
\frac{1}{2} (\Bar{n_L})^{\sss c} \,M\, n_L + h.c.
\end{eqnarray}
where
\begin{equation}
n_L \!\equiv\! \left( \begin{array}{c}
    \nu_L \\
    (\nu_L)^{\sss c}
    \end{array} \right) \nn,\nn
M \!\equiv\! \left( \begin{array}{cc}
    M_L & M^D_\nu\\
    M^D_\nu & M_R
    \end{array} \right) \nn;
\end{equation}
\noindent
$M^D_\nu$ is the standard $SU(2)\times U(1)$ breaking Dirac mass
term, and $M_L$ and $M_R$ are the isosinglet Majorana mass
terms for left-handed and right-handed neutrinos, respectively.

\smallskip

Supposing that the left-handed Majorana mass $M_L$ terms are
comparatively negligible, because of SM gauge symmetry protection,
a naturally small effective Majorana mass for the light
neutrinos (predominantly $\nu_{\sss L}$) can be generated by mixing
with the heavy states (predominantly $\nu_{\sss R}$) of mass
$M_{\nu_{\sss R}}$.
With no left-left term, $M_L=0$, the light eigenvalues
of the matrix $M$ are
\begin{equation}
  \label{eq:meff}
  M_{\rm eff} \! \approx \! M^D_\nu\,M_R^{-1}\,(M^D_\nu)^T\nn.
\end{equation}

This result is the well-known see-saw mechanism
\cite{seesaw}: the
light neutrino masses are quadratic in the Dirac masses and
inversely proportional to the large $\nu_R$ Majorana masses. Notice that
if some $\nu_{\sss R}$ are massless or light they would not
be integrated away but simply added to the light neutrinos.

\bigskip

We have already given the quantum charges of the Higgs fields,
$S$, $W$, $T$, $\xi$, $\phi_{\sss WS}$, $\phi_{\sss B-L}$ and
$\chi$ in Table $1$. With this quantum number choice of Higgs
fields the mass matrices are given for
\noindent%
the uct-quarks:%
\begin{equation}
M_U \simeq \frac{\sVEV{\phi_{\sss\rm WS}}}{\sqrt{2}}\hspace{-0.1cm}
\left ( \begin{array}{ccc}
        S^{\dagger}W^{\dagger}T^2(\xi^{\dagger})^2
        & W^{\dagger}T^2\xi & (W^{\dagger})^2T\xi \\
        S^{\dagger}W^{\dagger}T^2(\xi^{\dagger})^3
        & W^{\dagger}T^2 & (W^{\dagger})^2T \\
        S^{\dagger}(\xi^{\dagger})^3 & 1 & W^{\dagger}T^{\dagger}
                        \end{array} \right ) \label{M_U}
\end{equation}\noindent %
the dsb-quarks:
\begin{equation}
M_D \simeq \frac{\sVEV{\phi_{\sss\rm WS}}}{\sqrt{2}}\hspace{-0.1cm}
\left ( \begin{array}{ccc}
        SW(T^{\dagger})^2\xi^2 & W(T^{\dagger})^2\xi & T^3\xi \\
        SW(T^{\dagger})^2\xi & W(T^{\dagger})^2 & T^3 \\
        SW^2(T^{\dagger})^4\xi & W^2(T^{\dagger})^4 & WT
                        \end{array} \right ) \label{M_D}
\end{equation}\noindent %
the charged leptons:
\begin{equation}
M_E \simeq \frac{\sVEV{\phi_{\sss\rm WS}}}{\sqrt{2}}\hspace{-0.1cm}
\left ( \hspace{-0.2 cm}\begin{array}{ccc}
        SW(T^{\dagger})^2\xi^2 & W(T^{\dagger})^2(\xi^{\dagger})^3
        & WT^4(\xi^{\dagger})^3\chi\\
        SW(T^{\dagger})^2\xi^5 & W(T^{\dagger})^2 &
        WT^4\chi\\
        S(W^{\dagger})^2T^4\xi^5 & (W^{\dagger})^2T^4 & WT
                        \end{array} \hspace{-0.2 cm}\right ) \label{M_E}
\end{equation}\noindent%
the Dirac neutrinos:
\begin{equation}
M^D_\nu \simeq \frac{\sVEV{\phi_{\sss\rm WS}}}{\sqrt{2}}\hspace{-0.1cm}
\left ( \hspace{-0.2 cm}\begin{array}{ccc}
        S^{\dagger}W^{\dagger}T^2(\xi^{\dagger})^2 & %
        W^{\dagger}T^2(\xi^{\dagger})^3
        & (W^\dagger)T^2(\xi^\dagger)^3\chi\\
        S^{\dagger}W^{\dagger}T^2\xi & W^{\dagger}T^2 &
        (W^\dagger)T^2\chi\\
        S^{\dagger}W^{\dagger}T^\dagger\xi\chi^\dagger&
        W^{\dagger}T^\dagger\chi^\dagger
        & W^{\dagger}T^{\dagger}
                        \end{array} \hspace{-0.2 cm}\right ) \label{Md_N}
\end{equation}\noindent %
and the Majorana neutrinos:
\begin{equation}
M_R \simeq  \sVEV{\phi_{\sss\rm B-L}}\hspace{-0.1cm}
\left (\hspace{-0.2 cm}\begin{array}{ccc}
S^\dagger\chi^\dagger\xi &  \chi^\dagger & 1 \\
 \chi^\dagger & S\chi^\dagger\xi^\dagger & S\xi^\dagger \\
 1 & S\xi^\dagger & S\chi\xi^\dagger
\end{array} \hspace{-0.2 cm}\right ) \label{Mr_N}
\end{equation}

Note that the random complex order of unity and factorial factors,
which are supposed to multiply all the mass matrix elements,
are not represented here. We will discuss these factors in
section \ref{sec:berechnung}.

\section{Charged fermion spectrum}
The mass matrices for the quarks, $M_{U}$ and $M_{D}$, happen
not to have been changed at all by the introduction of the
``new'' Higgs fields $\chi$ (and $\phi_{\sss B-L}$, but
that has so little VEV compared to the Planck scale that it could
never compete). Even in the charged lepton mass matrix
the appearance of $\chi$ only occurs on off-diagonal matrix
elements. These elements are already small and remain so small as to
have no significance for the charged lepton mass predictions,
as long as $\chi$ is of the order $\sVEV{\chi}\approx0.07$ as we need
for fitting $\theta_{\rm atm}$.

Therefore all the fits of the ``old'' Anti-GUT model are valid and we
can still use the parameter values obtained by these earlier fits to
$S$, $W$, $T$, $\xi$, presented above in equation (\ref{eq:vevs}).
The result of a more recent fit \cite{fnsnew} to the charged fermion
spectrum, without imposing the constraint $\sVEV{S}=1$,
is shown in Table \ref{charged}.
\begin{table}
\caption{Typical fit, $\alpha=-1$, $\beta=1$, $\gamma=1$, $\delta=1$.}
\begin{tabular}{|c|c|c|} \hline
& Fitted &``Experiment''\\ \hline
$m_u$& $3.1~\MeV$&  $4~\MeV$\\
$m_d$& $6.6~\MeV$&  $9~\MeV$\\
$m_e$& $0.76~\MeV$& $0.5~\MeV$\\
\hline
$m_c$ & $1.29~\GeV$ & $1.4~\GeV$\\
$m_s$ & $390~\MeV$ &   $200~\MeV$\\
$m_{\mu}$ & $85~\MeV$ & $105~\MeV$\\
\hline
$M_t$ & $179~\GeV$ & $180~\GeV$\\
$m_b$ &  $7.8~\GeV$ & $6.3~\GeV$\\
$m_{\tau}$ & $1.29~\GeV$ & $1.78~\GeV$\\
\hline
$V_{us}$ & $0.21$   &  $0.22$\\
$V_{cb}$ & $ 0.023$ & $0.041$\\
$V_{ub}$ & $0.0050$ & $0.0035$\\
\hline
$J_{CP}$ & $1.04\cdot 10^{-5}$ & $(2\;-\;3.5)\cdot 10^{-5}$\\
\hline
``$\chi^2$'' &$1.46$&\\
\hline
\end{tabular}
\label{charged}
\end{table}

The quark and charged lepton fit of Table \ref{charged}
is for the Higgs charge assignments chosen according to a
set of discrete parameters
\begin{equation}
 \alpha=1,\;\; \beta=1,\;\; \gamma=-1,\;\; \delta=1.
\label{discrete}
\end{equation}
However, since the
effect of this choice only comes in via a correction which we call
the ``factorial correction'' \cite{fnsnew} and only makes changes of order
unity in principle, these discrete parameters should not be counted as
variable parameters in the fit---in fact we only allowed them
to have the values -1, 0, 1.
The Higgs VEV parameters for this choice (\ref{discrete}),
with the inclusion of the ``factorial correction'', are
\begin{equation}
\sVEV{W}=0.0894,\;\; \sVEV{T}=0.0525,\;\; \sVEV{S}=0.756,\;\;
\sVEV{\xi}= 0.0247.
\end{equation}

It is seen that it is a rather good fit to the data,
from the point of view that
it is only expected to work up to order of unity factors. The worst
case is the strange quark mass, since the expression for the
Jarlskog-triangle-area, which is a measure of CP-violation
$J_{CP}$, contains so many factors that we expect
the uncertainty in our prediction for it to be relatively large.
The trouble
with the strange quark is due to the fact
that the model has an order of magnitude
family degeneracy built into the diagonal elements of the
mass matrices. Thus, apart from
the charm and the top quarks whose masses are dominated
by off-diagonal mass matrix elements, the charged fermion masses
are predicted to have an order of magnitude family degeneracy.
So we cannot avoid having~$\eg$~the $SU(5)$ relations
(\ref{e4}) order-of-magnitudewise. Still having eq.~(\ref{e4}) only
as an order of magnitude relation is much better than getting it
as an exact prediction!

In fact we can even predict~\cite{fnsnew} the accuracy with which
we expect our predictions to work---namely that the uncertainty in the
logarithm shall be equal to that for the distribution of the logarithm of
a Gaussian distribution. Our fits actually do rather a bit too
well, and even agree with the prediction concerning the skewness of the
distribution:
that the worst deviation should be that some mass(es) turn out to be too
\underline{small} experimentally compared to the prediction---the
deviation of the strange quark mass fits in this way.

\section{Calculation of $M_{\rm eff}$}
\label{sec:berechnung}

In this section we calculate the effective neutrino mass matrix for
left-handed components. Since, strictly speaking, our model only
predicts orders of magnitude, a crude calculation is in principle
justified. This calculation is presented in the first
subsection, and then in the next subsection we
make ``statistical calculations'' with random order-one
factors and ``factorial factors''.

\smallskip

\subsection{Crude calculation}

 From equation~(\ref{Mr_N}) we see, to the first approximation,
that there are one massless and two degenerate right-handed
neutrinos coming from the VEV of the $B\!-\!L$ breaking
Higgs field, $\sVEV{\phi_{\sss\rm B-L}}$.
The mass splitting between the two almost degenerate see-saw
neutrinos is $M_{31}\sVEV{S}\sVEV{\chi}\sVEV{\xi}$,
where $M_{31}\approx\sVEV{\phi_{\sss\rm B-L}}$ is the approximately
common mass of the two heaviest see-saw neutrinos. The third lightest
see-saw neutrino is dominantly ``proto second generation''
and has the mass $\sVEV{\phi_{\sss\rm B-L}}\sVEV{\chi}\sVEV{\xi}$.

For the left-handed neutrinos, to the first approximation, we get
the effective mass matrix as follows:
\begin{equation}
M_{\rm eff}\approx
\frac{W^2T^2\sVEV{\phi_{\sss WS}}^2}{2\sVEV{\phi_{\sss\rm B-L}}}
\left( \begin{array}{ccc}
\frac{T^2\xi^5}{\chi} & \frac{T^2\xi^2}{\chi} & T\xi^2 \\
\frac{T^2\xi^2}{\chi} & \frac{T}{\xi} & \frac{T}{\xi}\\
T\xi^2 &   \frac{T}{\xi} & \frac{\chi}{\xi} \\
\end{array} \right) \nn,
\label{eq:hmatrix}
\end{equation}

But we have to emphasise here that this approximation
is {\em good enough to calculate only the heaviest left-handed neutrino}.
This is because all the mass matrix elements, to this approximation,
come from the propagator contribution of the lightest see-saw
particle, so that they really form a degenerate matrix of
rank one.  Using this contribution only would lead to two left-handed
massless neutrinos and one massive left-handed neutrino.
But we can still obtain the
mixing angles $\theta_{13}$ and $\theta_{23}$ and the heaviest
mass from
$M_{\rm eff}$:
\begin{eqnarray}
  \label{eq:mixing1223}
  \theta_{13}&=&\theta_{e{},{\rm heavy}}\approx\frac{T}{\chi} \xi^3 \\
  \theta_{23}&=&\theta_{\mu{},{\rm heavy}}\approx
\left\{ \begin{array}{r@{\hspace{1cm}}l}
       \frac{T}{\chi} & \mbox{when $\chi\sgeq T$} \\
       1              & \mbox{when $\chi\sleq T$} \\
\end{array} \right. \\
M_{\sss \nu_{\sss L}{\rm heavy}}\!\!\! &\approx&
\left\{ \begin{array}{r@{\hspace{1cm}}l}
\frac{W^2T^2\sVEV{\phi_{\sss WS}}^2}{2\sVEV{\phi_{\sss\rm B-L}}}
\frac{\chi}{\xi} & \mbox{when $\chi\sgeq T$} \\
\frac{W^2T^2\sVEV{\phi_{\sss WS}}^2}{2\sVEV{\phi_{\sss\rm B-L}}}
\frac{T}{\xi} & \mbox{when $\chi\sleq T$} \\
\end{array} \right. \end{eqnarray}
{}From these equations we can restrict the region of $\chi$ by
comparing with Super-Kamiokande experimental data; $\chi$ must be
almost of the same order as $T$. Thus we know the mixing angle
$\theta_{13}$ of
the first and third generations must be of the order of $\xi^3$.

\bigskip
\indent
However, to get the much lower neutrino masses we cannot use the
contribution from the lightest see-saw propagator, but we have
to use the propagator terms from the two approximately equally
heavy see-saw particles. This contribution to the propagator
matrix is
\begin{equation}
  \label{eq:invmr}
M^{-1}_{{\sss R}}{}_{\big|}{}_{{\tiny\rm{{\begin{array}{l} heavy\\
see-saws\end{array}}}}}%
\!\!\!\!\approx \frac{1}{\sVEV{\phi_{\sss\rm B-L}}}\hspace{-0.1cm}
\left (\hspace{-0.2 cm}\begin{array}{ccc}
\chi\xi &  \xi & 1 \\
 \xi& \chi\xi & \chi \\
 1 & \chi & \chi\xi
\end{array} \hspace{-0.2 cm}\right )
\end{equation}
where the $\chi\xi/\sVEV{\phi_{\sss B-L}}$ comes from the mass difference
of the almost degenerate see-saw particles.

Surrounding this propagator contribution with the ``Dirac $\nu$''-mass
matrix we get
\begin{eqnarray}
  \label{eq:appsee}
M_{{\rm eff}}{}_{\big|}{}_{{\tiny\rm{{\begin{array}{l} heavy\\
see-saws\end{array}}}}}%
\!\!\!\!&\approx&  M^D_\nu\,M^{-1}_{{\sss R}}{}_{\big|}
{}_{{\tiny\rm{{\begin{array}{l} heavy\\ see-saws\end{array}}}}}%
\!\!(M^D_\nu)^T\nonumber\\
 &\approx&  \frac{W^2T^2\sVEV{\phi_{\sss WS}}^2}
{2\sVEV{\phi_{\sss\rm B-L}}}\hspace
{-0.1cm}
\left (\hspace{-0.2 cm}\begin{array}{ccc}
T^2\xi^6 &  T\xi^3 & T\xi^2 \\
T\xi^3 & T^2\chi\xi & T\chi \\
 T\xi^2 & T\chi & \chi\xi
\end{array} \hspace{-0.2 cm}\right )\nn.
\end{eqnarray}
It is from this contribution that the two lightest left-handed neutrino
masses and their mixing angle, $\theta_{12}$, are obtained:
\begin{eqnarray}
  &&M_{\sss \nu_{\sss L}{\rm medium}}\approx
\frac{W^2T^2\chi\xi\sVEV{\phi_{\sss WS}}^2}%
{2\sVEV{\phi_{\sss\rm B-L}}} \\
  &&\theta_{12}=\theta_{e{},{\rm medium}}\approx
\left\{ \begin{array}{r@{\hspace{1cm}}l}
       \frac{T}{\chi}\xi & \mbox{when $\chi\sgeq T$} \\
       \xi               & \mbox{when $\chi\sleq T$}\nn. \\
\end{array} \right.
\end{eqnarray}
Note that the lightest mass particle is dominantly the $\nu_{e_{\sss L}}$
neutrino
and its small mixing is mainly with the medium mass neutrino (%
$\theta_{e{},{\rm medium}}/\theta_{e{},{\rm heavy}}\approx\xi^{-2}\gg 1$).
So we should approximately
identify the solar oscillation mixing angle with
the mixing to the medium heavy neutrino:
\begin{equation}
  \label{eq:solxi}
  \theta_{\odot}\simeq \theta_{e{},{\rm medium}}\approx \xi
\end{equation}%
and the solar mass squared difference as:
\begin{equation}
  \label{eq:solms}
  \Delta m_{\odot }^2 \approx M_{\sss \nu_{\sss L}{\rm medium}}^2\approx
\frac{W^4T^4\chi^2\xi^2\sVEV{\phi_{\sss WS}}^4}%
{4\sVEV{\phi_{\sss\rm B-L}}^2} \nn.
\end{equation}
The atmospheric mixing angle goes between the heaviest and the medium
mass neutrino:
\begin{eqnarray}
  \label{eq:atmms}
  \Delta m_{\rm atm}^2 &\approx& M_{\sss \nu_{\sss L}{\rm heavy}}^2-%
M_{\sss \nu_{\sss L}{\rm medium}}^2 \nonumber\\
&\approx& \frac{W^4T^4\chi^2\sVEV{\phi_{\sss WS}}^4}%
{4\sVEV{\phi_{\sss\rm B-L}}^2\xi^2}
\end{eqnarray}
{}From equations (\ref{eq:solms}) and (\ref{eq:atmms}), we find that the ratio
of solar and atmospheric neutrino mass squared differences must be of the
order of $\xi^4$, say about $10^{-4}$.

\subsection{Statistical calculation using random order unity factors}
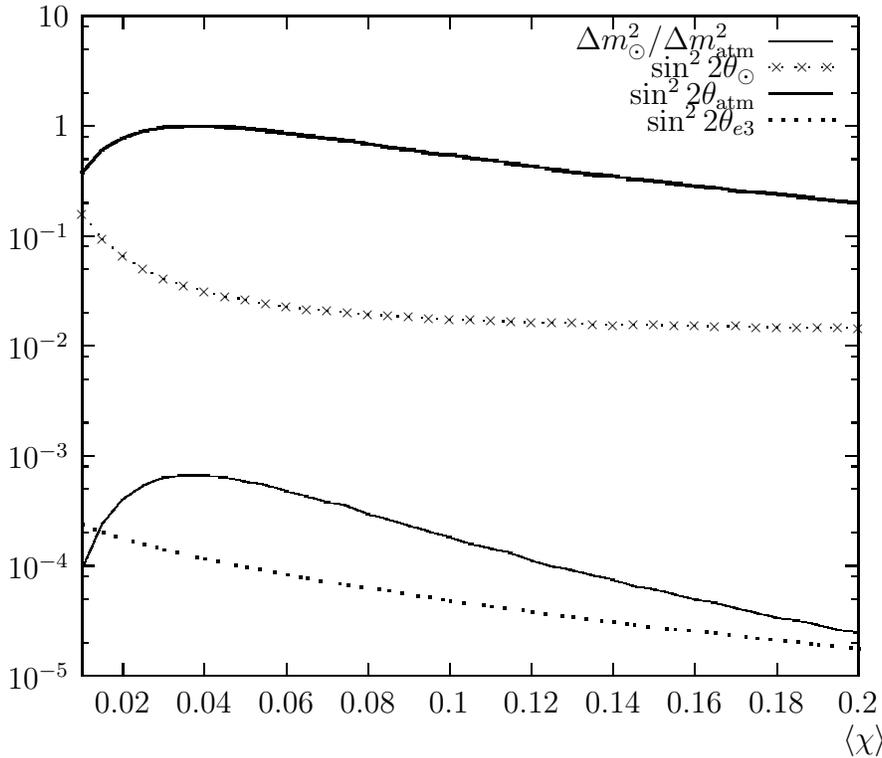
\begin{figure}[ht!]
\begin{center}
\label{fig:resuneu}
\setlength{\unitlength}{0.240900pt}
\ifx\plotpoint\undefined\newsavebox{\plotpoint}\fi
\sbox{\plotpoint}{\rule[-0.200pt]{0.400pt}{0.400pt}}%
\begin{picture}(1500,1200)(0,0)
\font\gnuplot=cmr10 at 10pt
\gnuplot
\sbox{\plotpoint}{\rule[-0.200pt]{0.400pt}{0.400pt}}%
\put(220.0,123.0){\rule[-0.200pt]{4.818pt}{0.400pt}}
\put(200,123){\makebox(0,0)[r]{$10^{-5}$}}
\put(1419.0,123.0){\rule[-0.200pt]{4.818pt}{0.400pt}}
\put(220.0,175.0){\rule[-0.200pt]{2.409pt}{0.400pt}}
\put(1429.0,175.0){\rule[-0.200pt]{2.409pt}{0.400pt}}
\put(220.0,244.0){\rule[-0.200pt]{2.409pt}{0.400pt}}
\put(1429.0,244.0){\rule[-0.200pt]{2.409pt}{0.400pt}}
\put(220.0,279.0){\rule[-0.200pt]{2.409pt}{0.400pt}}
\put(1429.0,279.0){\rule[-0.200pt]{2.409pt}{0.400pt}}
\put(220.0,296.0){\rule[-0.200pt]{4.818pt}{0.400pt}}
\put(200,296){\makebox(0,0)[r]{$10^{-4}$}}
\put(1419.0,296.0){\rule[-0.200pt]{4.818pt}{0.400pt}}
\put(220.0,348.0){\rule[-0.200pt]{2.409pt}{0.400pt}}
\put(1429.0,348.0){\rule[-0.200pt]{2.409pt}{0.400pt}}
\put(220.0,417.0){\rule[-0.200pt]{2.409pt}{0.400pt}}
\put(1429.0,417.0){\rule[-0.200pt]{2.409pt}{0.400pt}}
\put(220.0,452.0){\rule[-0.200pt]{2.409pt}{0.400pt}}
\put(1429.0,452.0){\rule[-0.200pt]{2.409pt}{0.400pt}}
\put(220.0,469.0){\rule[-0.200pt]{4.818pt}{0.400pt}}
\put(200,469){\makebox(0,0)[r]{$10^{-3}$}}
\put(1419.0,469.0){\rule[-0.200pt]{4.818pt}{0.400pt}}
\put(220.0,521.0){\rule[-0.200pt]{2.409pt}{0.400pt}}
\put(1429.0,521.0){\rule[-0.200pt]{2.409pt}{0.400pt}}
\put(220.0,589.0){\rule[-0.200pt]{2.409pt}{0.400pt}}
\put(1429.0,589.0){\rule[-0.200pt]{2.409pt}{0.400pt}}
\put(220.0,625.0){\rule[-0.200pt]{2.409pt}{0.400pt}}
\put(1429.0,625.0){\rule[-0.200pt]{2.409pt}{0.400pt}}
\put(220.0,642.0){\rule[-0.200pt]{4.818pt}{0.400pt}}
\put(200,642){\makebox(0,0)[r]{$10^{-2}$}}
\put(1419.0,642.0){\rule[-0.200pt]{4.818pt}{0.400pt}}
\put(220.0,694.0){\rule[-0.200pt]{2.409pt}{0.400pt}}
\put(1429.0,694.0){\rule[-0.200pt]{2.409pt}{0.400pt}}
\put(220.0,762.0){\rule[-0.200pt]{2.409pt}{0.400pt}}
\put(1429.0,762.0){\rule[-0.200pt]{2.409pt}{0.400pt}}
\put(220.0,798.0){\rule[-0.200pt]{2.409pt}{0.400pt}}
\put(1429.0,798.0){\rule[-0.200pt]{2.409pt}{0.400pt}}
\put(220.0,814.0){\rule[-0.200pt]{4.818pt}{0.400pt}}
\put(200,814){\makebox(0,0)[r]{$10^{-1}$}}
\put(1419.0,814.0){\rule[-0.200pt]{4.818pt}{0.400pt}}
\put(220.0,866.0){\rule[-0.200pt]{2.409pt}{0.400pt}}
\put(1429.0,866.0){\rule[-0.200pt]{2.409pt}{0.400pt}}
\put(220.0,935.0){\rule[-0.200pt]{2.409pt}{0.400pt}}
\put(1429.0,935.0){\rule[-0.200pt]{2.409pt}{0.400pt}}
\put(220.0,970.0){\rule[-0.200pt]{2.409pt}{0.400pt}}
\put(1429.0,970.0){\rule[-0.200pt]{2.409pt}{0.400pt}}
\put(220.0,987.0){\rule[-0.200pt]{4.818pt}{0.400pt}}
\put(200,987){\makebox(0,0)[r]{$1$}}
\put(1419.0,987.0){\rule[-0.200pt]{4.818pt}{0.400pt}}
\put(220.0,1039.0){\rule[-0.200pt]{2.409pt}{0.400pt}}
\put(1429.0,1039.0){\rule[-0.200pt]{2.409pt}{0.400pt}}
\put(220.0,1108.0){\rule[-0.200pt]{2.409pt}{0.400pt}}
\put(1429.0,1108.0){\rule[-0.200pt]{2.409pt}{0.400pt}}
\put(220.0,1143.0){\rule[-0.200pt]{2.409pt}{0.400pt}}
\put(1429.0,1143.0){\rule[-0.200pt]{2.409pt}{0.400pt}}
\put(220.0,1160.0){\rule[-0.200pt]{4.818pt}{0.400pt}}
\put(200,1160){\makebox(0,0)[r]{$10$}}
\put(1419.0,1160.0){\rule[-0.200pt]{4.818pt}{0.400pt}}
\put(284.0,123.0){\rule[-0.200pt]{0.400pt}{4.818pt}}
\put(284,82){\makebox(0,0){$0.02$}}
\put(284.0,1140.0){\rule[-0.200pt]{0.400pt}{4.818pt}}
\put(412.0,123.0){\rule[-0.200pt]{0.400pt}{4.818pt}}
\put(412,82){\makebox(0,0){$0.04$}}
\put(412.0,1140.0){\rule[-0.200pt]{0.400pt}{4.818pt}}
\put(541.0,123.0){\rule[-0.200pt]{0.400pt}{4.818pt}}
\put(541,82){\makebox(0,0){$0.06$}}
\put(541.0,1140.0){\rule[-0.200pt]{0.400pt}{4.818pt}}
\put(669.0,123.0){\rule[-0.200pt]{0.400pt}{4.818pt}}
\put(669,82){\makebox(0,0){$0.08$}}
\put(669.0,1140.0){\rule[-0.200pt]{0.400pt}{4.818pt}}
\put(797.0,123.0){\rule[-0.200pt]{0.400pt}{4.818pt}}
\put(797,82){\makebox(0,0){$0.1$}}
\put(797.0,1140.0){\rule[-0.200pt]{0.400pt}{4.818pt}}
\put(926.0,123.0){\rule[-0.200pt]{0.400pt}{4.818pt}}
\put(926,82){\makebox(0,0){$0.12$}}
\put(926.0,1140.0){\rule[-0.200pt]{0.400pt}{4.818pt}}
\put(1054.0,123.0){\rule[-0.200pt]{0.400pt}{4.818pt}}
\put(1054,82){\makebox(0,0){$0.14$}}
\put(1054.0,1140.0){\rule[-0.200pt]{0.400pt}{4.818pt}}
\put(1182.0,123.0){\rule[-0.200pt]{0.400pt}{4.818pt}}
\put(1182,82){\makebox(0,0){$0.16$}}
\put(1182.0,1140.0){\rule[-0.200pt]{0.400pt}{4.818pt}}
\put(1311.0,123.0){\rule[-0.200pt]{0.400pt}{4.818pt}}
\put(1311,82){\makebox(0,0){$0.18$}}
\put(1311.0,1140.0){\rule[-0.200pt]{0.400pt}{4.818pt}}
\put(1439.0,123.0){\rule[-0.200pt]{0.400pt}{4.818pt}}
\put(1439,82){\makebox(0,0){$0.2$}}
\put(1439.0,1140.0){\rule[-0.200pt]{0.400pt}{4.818pt}}
\put(220.0,123.0){\rule[-0.200pt]{293.657pt}{0.400pt}}
\put(1439.0,123.0){\rule[-0.200pt]{0.400pt}{249.813pt}}
\put(220.0,1160.0){\rule[-0.200pt]{293.657pt}{0.400pt}}
\put(1450,21){\makebox(0,0){$\sVEV{\chi}$}}
\put(220.0,123.0){\rule[-0.200pt]{0.400pt}{249.813pt}}
\put(1279,1120){\makebox(0,0)[r]{${
\footnotesize\Delta m^2_{\odot}/\Delta m^2_{\rm atm}}$}}
\put(1299.0,1120.0){\rule[-0.200pt]{24.090pt}{0.400pt}}
\put(220,289){\usebox{\plotpoint}}
\multiput(220.58,289.00)(0.497,1.131){61}{\rule{0.120pt}{1.000pt}}
\multiput(219.17,289.00)(32.000,69.924){2}{\rule{0.400pt}{0.500pt}}
\multiput(252.58,361.00)(0.497,0.609){61}{\rule{0.120pt}{0.588pt}}
\multiput(251.17,361.00)(32.000,37.781){2}{\rule{0.400pt}{0.294pt}}
\multiput(284.00,400.58)(0.729,0.496){41}{\rule{0.682pt}{0.120pt}}
\multiput(284.00,399.17)(30.585,22.000){2}{\rule{0.341pt}{0.400pt}}
\multiput(316.00,422.58)(1.250,0.493){23}{\rule{1.085pt}{0.119pt}}
\multiput(316.00,421.17)(29.749,13.000){2}{\rule{0.542pt}{0.400pt}}
\multiput(348.00,435.61)(6.937,0.447){3}{\rule{4.367pt}{0.108pt}}
\multiput(348.00,434.17)(22.937,3.000){2}{\rule{2.183pt}{0.400pt}}
\multiput(412.00,436.95)(7.160,-0.447){3}{\rule{4.500pt}{0.108pt}}
\multiput(412.00,437.17)(23.660,-3.000){2}{\rule{2.250pt}{0.400pt}}
\multiput(445.00,433.93)(2.399,-0.485){11}{\rule{1.929pt}{0.117pt}}
\multiput(445.00,434.17)(27.997,-7.000){2}{\rule{0.964pt}{0.400pt}}
\multiput(477.00,426.93)(3.493,-0.477){7}{\rule{2.660pt}{0.115pt}}
\multiput(477.00,427.17)(26.479,-5.000){2}{\rule{1.330pt}{0.400pt}}
\multiput(509.00,421.92)(1.642,-0.491){17}{\rule{1.380pt}{0.118pt}}
\multiput(509.00,422.17)(29.136,-10.000){2}{\rule{0.690pt}{0.400pt}}
\multiput(541.00,411.93)(2.079,-0.488){13}{\rule{1.700pt}{0.117pt}}
\multiput(541.00,412.17)(28.472,-8.000){2}{\rule{0.850pt}{0.400pt}}
\multiput(573.00,403.93)(1.834,-0.489){15}{\rule{1.522pt}{0.118pt}}
\multiput(573.00,404.17)(28.841,-9.000){2}{\rule{0.761pt}{0.400pt}}
\multiput(605.00,394.93)(2.841,-0.482){9}{\rule{2.233pt}{0.116pt}}
\multiput(605.00,395.17)(27.365,-6.000){2}{\rule{1.117pt}{0.400pt}}
\multiput(637.00,388.92)(1.250,-0.493){23}{\rule{1.085pt}{0.119pt}}
\multiput(637.00,389.17)(29.749,-13.000){2}{\rule{0.542pt}{0.400pt}}
\multiput(669.00,375.93)(2.079,-0.488){13}{\rule{1.700pt}{0.117pt}}
\multiput(669.00,376.17)(28.472,-8.000){2}{\rule{0.850pt}{0.400pt}}
\multiput(701.00,367.92)(1.642,-0.491){17}{\rule{1.380pt}{0.118pt}}
\multiput(701.00,368.17)(29.136,-10.000){2}{\rule{0.690pt}{0.400pt}}
\multiput(733.00,357.93)(1.834,-0.489){15}{\rule{1.522pt}{0.118pt}}
\multiput(733.00,358.17)(28.841,-9.000){2}{\rule{0.761pt}{0.400pt}}
\multiput(765.00,348.93)(1.834,-0.489){15}{\rule{1.522pt}{0.118pt}}
\multiput(765.00,349.17)(28.841,-9.000){2}{\rule{0.761pt}{0.400pt}}
\multiput(797.00,339.92)(1.534,-0.492){19}{\rule{1.300pt}{0.118pt}}
\multiput(797.00,340.17)(30.302,-11.000){2}{\rule{0.650pt}{0.400pt}}
\multiput(830.00,328.93)(2.399,-0.485){11}{\rule{1.929pt}{0.117pt}}
\multiput(830.00,329.17)(27.997,-7.000){2}{\rule{0.964pt}{0.400pt}}
\multiput(862.00,321.93)(2.399,-0.485){11}{\rule{1.929pt}{0.117pt}}
\multiput(862.00,322.17)(27.997,-7.000){2}{\rule{0.964pt}{0.400pt}}
\multiput(894.00,314.92)(1.358,-0.492){21}{\rule{1.167pt}{0.119pt}}
\multiput(894.00,315.17)(29.579,-12.000){2}{\rule{0.583pt}{0.400pt}}
\multiput(926.00,302.93)(1.834,-0.489){15}{\rule{1.522pt}{0.118pt}}
\multiput(926.00,303.17)(28.841,-9.000){2}{\rule{0.761pt}{0.400pt}}
\multiput(958.00,293.93)(2.841,-0.482){9}{\rule{2.233pt}{0.116pt}}
\multiput(958.00,294.17)(27.365,-6.000){2}{\rule{1.117pt}{0.400pt}}
\multiput(990.00,287.93)(2.079,-0.488){13}{\rule{1.700pt}{0.117pt}}
\multiput(990.00,288.17)(28.472,-8.000){2}{\rule{0.850pt}{0.400pt}}
\multiput(1022.00,279.93)(2.399,-0.485){11}{\rule{1.929pt}{0.117pt}}
\multiput(1022.00,280.17)(27.997,-7.000){2}{\rule{0.964pt}{0.400pt}}
\multiput(1054.00,272.92)(1.642,-0.491){17}{\rule{1.380pt}{0.118pt}}
\multiput(1054.00,273.17)(29.136,-10.000){2}{\rule{0.690pt}{0.400pt}}
\multiput(1086.00,262.93)(3.493,-0.477){7}{\rule{2.660pt}{0.115pt}}
\multiput(1086.00,263.17)(26.479,-5.000){2}{\rule{1.330pt}{0.400pt}}
\multiput(1118.00,257.93)(2.079,-0.488){13}{\rule{1.700pt}{0.117pt}}
\multiput(1118.00,258.17)(28.472,-8.000){2}{\rule{0.850pt}{0.400pt}}
\multiput(1150.00,249.93)(2.079,-0.488){13}{\rule{1.700pt}{0.117pt}}
\multiput(1150.00,250.17)(28.472,-8.000){2}{\rule{0.850pt}{0.400pt}}
\multiput(1182.00,241.93)(3.493,-0.477){7}{\rule{2.660pt}{0.115pt}}
\multiput(1182.00,242.17)(26.479,-5.000){2}{\rule{1.330pt}{0.400pt}}
\multiput(1214.00,236.93)(1.893,-0.489){15}{\rule{1.567pt}{0.118pt}}
\multiput(1214.00,237.17)(29.748,-9.000){2}{\rule{0.783pt}{0.400pt}}
\multiput(1247.00,227.93)(2.399,-0.485){11}{\rule{1.929pt}{0.117pt}}
\multiput(1247.00,228.17)(27.997,-7.000){2}{\rule{0.964pt}{0.400pt}}
\multiput(1279.00,220.93)(2.079,-0.488){13}{\rule{1.700pt}{0.117pt}}
\multiput(1279.00,221.17)(28.472,-8.000){2}{\rule{0.850pt}{0.400pt}}
\multiput(1311.00,212.94)(4.575,-0.468){5}{\rule{3.300pt}{0.113pt}}
\multiput(1311.00,213.17)(25.151,-4.000){2}{\rule{1.650pt}{0.400pt}}
\multiput(1343.00,208.93)(2.399,-0.485){11}{\rule{1.929pt}{0.117pt}}
\multiput(1343.00,209.17)(27.997,-7.000){2}{\rule{0.964pt}{0.400pt}}
\multiput(1375.00,201.93)(2.079,-0.488){13}{\rule{1.700pt}{0.117pt}}
\multiput(1375.00,202.17)(28.472,-8.000){2}{\rule{0.850pt}{0.400pt}}
\multiput(1407.00,193.94)(4.575,-0.468){5}{\rule{3.300pt}{0.113pt}}
\multiput(1407.00,194.17)(25.151,-4.000){2}{\rule{1.650pt}{0.400pt}}
\put(380.0,438.0){\rule[-0.200pt]{7.709pt}{0.400pt}}
\put(1279,1079){\makebox(0,0)[r]{${\footnotesize\sin^22\theta_{\odot}}$}}
\multiput(1299,1079)(20.756,0.000){5}{\usebox{\plotpoint}}
\put(1399,1079){\usebox{\plotpoint}}
\put(220,849){\usebox{\plotpoint}}
\multiput(220,849)(12.966,-16.207){3}{\usebox{\plotpoint}}
\multiput(252,809)(16.109,-13.088){2}{\usebox{\plotpoint}}
\multiput(284,783)(17.601,-11.000){2}{\usebox{\plotpoint}}
\put(329.50,756.25){\usebox{\plotpoint}}
\multiput(348,747)(19.628,-6.747){2}{\usebox{\plotpoint}}
\multiput(380,736)(19.980,-5.619){2}{\usebox{\plotpoint}}
\put(427.57,723.23){\usebox{\plotpoint}}
\multiput(445,719)(20.507,-3.204){2}{\usebox{\plotpoint}}
\multiput(477,714)(20.276,-4.435){2}{\usebox{\plotpoint}}
\put(529.53,704.43){\usebox{\plotpoint}}
\multiput(541,703)(20.595,-2.574){2}{\usebox{\plotpoint}}
\put(591.43,697.85){\usebox{\plotpoint}}
\multiput(605,697)(20.595,-2.574){2}{\usebox{\plotpoint}}
\put(653.39,691.98){\usebox{\plotpoint}}
\multiput(669,691)(20.715,-1.295){2}{\usebox{\plotpoint}}
\put(715.53,688.09){\usebox{\plotpoint}}
\multiput(733,687)(20.715,-1.295){2}{\usebox{\plotpoint}}
\put(777.68,684.21){\usebox{\plotpoint}}
\multiput(797,683)(20.746,-0.629){2}{\usebox{\plotpoint}}
\multiput(830,682)(20.745,-0.648){2}{\usebox{\plotpoint}}
\put(881.35,679.79){\usebox{\plotpoint}}
\multiput(894,679)(20.745,-0.648){2}{\usebox{\plotpoint}}
\put(943.57,678.00){\usebox{\plotpoint}}
\multiput(958,678)(20.756,0.000){2}{\usebox{\plotpoint}}
\put(1005.77,676.52){\usebox{\plotpoint}}
\multiput(1022,675)(20.745,-0.648){2}{\usebox{\plotpoint}}
\put(1067.94,674.44){\usebox{\plotpoint}}
\multiput(1086,675)(20.756,0.000){2}{\usebox{\plotpoint}}
\put(1130.18,674.24){\usebox{\plotpoint}}
\multiput(1150,673)(20.756,0.000){2}{\usebox{\plotpoint}}
\multiput(1182,673)(20.745,-0.648){2}{\usebox{\plotpoint}}
\put(1233.89,672.60){\usebox{\plotpoint}}
\multiput(1247,673)(20.715,-1.295){2}{\usebox{\plotpoint}}
\put(1296.09,671.00){\usebox{\plotpoint}}
\multiput(1311,671)(20.745,-0.648){2}{\usebox{\plotpoint}}
\put(1358.34,670.00){\usebox{\plotpoint}}
\multiput(1375,670)(20.745,0.648){2}{\usebox{\plotpoint}}
\put(1420.57,670.15){\usebox{\plotpoint}}
\put(1439,669){\usebox{\plotpoint}}
\put(220,849){\makebox(0,0){${\sss \times}$}}
\put(252,809){\makebox(0,0){${\sss \times}$}}
\put(284,783){\makebox(0,0){${\sss \times}$}}
\put(316,763){\makebox(0,0){${\sss \times}$}}
\put(348,747){\makebox(0,0){${\sss \times}$}}
\put(380,736){\makebox(0,0){${\sss \times}$}}
\put(412,727){\makebox(0,0){${\sss \times}$}}
\put(445,719){\makebox(0,0){${\sss \times}$}}
\put(477,714){\makebox(0,0){${\sss \times}$}}
\put(509,707){\makebox(0,0){${\sss \times}$}}
\put(541,703){\makebox(0,0){${\sss \times}$}}
\put(573,699){\makebox(0,0){${\sss \times}$}}
\put(605,697){\makebox(0,0){${\sss \times}$}}
\put(637,693){\makebox(0,0){${\sss \times}$}}
\put(669,691){\makebox(0,0){${\sss \times}$}}
\put(701,689){\makebox(0,0){${\sss \times}$}}
\put(733,687){\makebox(0,0){${\sss \times}$}}
\put(765,685){\makebox(0,0){${\sss \times}$}}
\put(797,683){\makebox(0,0){${\sss \times}$}}
\put(830,682){\makebox(0,0){${\sss \times}$}}
\put(862,681){\makebox(0,0){${\sss \times}$}}
\put(894,679){\makebox(0,0){${\sss \times}$}}
\put(926,678){\makebox(0,0){${\sss \times}$}}
\put(958,678){\makebox(0,0){${\sss \times}$}}
\put(990,678){\makebox(0,0){${\sss \times}$}}
\put(1022,675){\makebox(0,0){${\sss \times}$}}
\put(1054,674){\makebox(0,0){${\sss \times}$}}
\put(1086,675){\makebox(0,0){${\sss \times}$}}
\put(1118,675){\makebox(0,0){${\sss \times}$}}
\put(1150,673){\makebox(0,0){${\sss \times}$}}
\put(1182,673){\makebox(0,0){${\sss \times}$}}
\put(1214,672){\makebox(0,0){${\sss \times}$}}
\put(1247,673){\makebox(0,0){${\sss \times}$}}
\put(1279,671){\makebox(0,0){${\sss \times}$}}
\put(1311,671){\makebox(0,0){${\sss \times}$}}
\put(1343,670){\makebox(0,0){${\sss \times}$}}
\put(1375,670){\makebox(0,0){${\sss \times}$}}
\put(1407,671){\makebox(0,0){${\sss \times}$}}
\put(1439,669){\makebox(0,0){${\sss \times}$}}
\put(1311,1079){\makebox(0,0){${\sss \times}$}}
\put(1351,1079){\makebox(0,0){${\sss \times}$}}
\put(1391,1079){\makebox(0,0){${\sss \times}$}}
\sbox{\plotpoint}{\rule[-0.400pt]{0.800pt}{0.800pt}}%
\put(1279,1038){\makebox(0,0)[r]{${\footnotesize\sin^22\theta_{\rm atm}}$}}
\put(1299.0,1038.0){\rule[-0.400pt]{24.090pt}{0.800pt}}
\put(220,913){\usebox{\plotpoint}}
\multiput(221.41,913.00)(0.503,0.577){57}{\rule{0.121pt}{1.125pt}}
\multiput(218.34,913.00)(32.000,34.665){2}{\rule{0.800pt}{0.563pt}}
\multiput(252.00,951.41)(0.902,0.506){29}{\rule{1.622pt}{0.122pt}}
\multiput(252.00,948.34)(28.633,18.000){2}{\rule{0.811pt}{0.800pt}}
\multiput(284.00,969.40)(1.536,0.512){15}{\rule{2.527pt}{0.123pt}}
\multiput(284.00,966.34)(26.755,11.000){2}{\rule{1.264pt}{0.800pt}}
\multiput(316.00,980.39)(3.365,0.536){5}{\rule{4.467pt}{0.129pt}}
\multiput(316.00,977.34)(22.729,6.000){2}{\rule{2.233pt}{0.800pt}}
\put(348,984.34){\rule{7.709pt}{0.800pt}}
\multiput(348.00,983.34)(16.000,2.000){2}{\rule{3.854pt}{0.800pt}}
\put(412,984.84){\rule{7.950pt}{0.800pt}}
\multiput(412.00,985.34)(16.500,-1.000){2}{\rule{3.975pt}{0.800pt}}
\put(445,982.84){\rule{7.709pt}{0.800pt}}
\multiput(445.00,984.34)(16.000,-3.000){2}{\rule{3.854pt}{0.800pt}}
\put(477,979.84){\rule{7.709pt}{0.800pt}}
\multiput(477.00,981.34)(16.000,-3.000){2}{\rule{3.854pt}{0.800pt}}
\put(509,976.34){\rule{6.600pt}{0.800pt}}
\multiput(509.00,978.34)(18.301,-4.000){2}{\rule{3.300pt}{0.800pt}}
\put(541,972.34){\rule{6.600pt}{0.800pt}}
\multiput(541.00,974.34)(18.301,-4.000){2}{\rule{3.300pt}{0.800pt}}
\put(573,968.34){\rule{6.600pt}{0.800pt}}
\multiput(573.00,970.34)(18.301,-4.000){2}{\rule{3.300pt}{0.800pt}}
\put(605,964.34){\rule{6.600pt}{0.800pt}}
\multiput(605.00,966.34)(18.301,-4.000){2}{\rule{3.300pt}{0.800pt}}
\multiput(637.00,962.06)(4.958,-0.560){3}{\rule{5.320pt}{0.135pt}}
\multiput(637.00,962.34)(20.958,-5.000){2}{\rule{2.660pt}{0.800pt}}
\multiput(669.00,957.06)(4.958,-0.560){3}{\rule{5.320pt}{0.135pt}}
\multiput(669.00,957.34)(20.958,-5.000){2}{\rule{2.660pt}{0.800pt}}
\put(701,950.34){\rule{6.600pt}{0.800pt}}
\multiput(701.00,952.34)(18.301,-4.000){2}{\rule{3.300pt}{0.800pt}}
\multiput(733.00,948.06)(4.958,-0.560){3}{\rule{5.320pt}{0.135pt}}
\multiput(733.00,948.34)(20.958,-5.000){2}{\rule{2.660pt}{0.800pt}}
\put(765,941.84){\rule{7.709pt}{0.800pt}}
\multiput(765.00,943.34)(16.000,-3.000){2}{\rule{3.854pt}{0.800pt}}
\multiput(797.00,940.06)(5.126,-0.560){3}{\rule{5.480pt}{0.135pt}}
\multiput(797.00,940.34)(21.626,-5.000){2}{\rule{2.740pt}{0.800pt}}
\put(830,933.34){\rule{6.600pt}{0.800pt}}
\multiput(830.00,935.34)(18.301,-4.000){2}{\rule{3.300pt}{0.800pt}}
\multiput(862.00,931.06)(4.958,-0.560){3}{\rule{5.320pt}{0.135pt}}
\multiput(862.00,931.34)(20.958,-5.000){2}{\rule{2.660pt}{0.800pt}}
\put(894,924.34){\rule{6.600pt}{0.800pt}}
\multiput(894.00,926.34)(18.301,-4.000){2}{\rule{3.300pt}{0.800pt}}
\multiput(926.00,922.06)(4.958,-0.560){3}{\rule{5.320pt}{0.135pt}}
\multiput(926.00,922.34)(20.958,-5.000){2}{\rule{2.660pt}{0.800pt}}
\put(958,915.34){\rule{6.600pt}{0.800pt}}
\multiput(958.00,917.34)(18.301,-4.000){2}{\rule{3.300pt}{0.800pt}}
\put(990,911.34){\rule{6.600pt}{0.800pt}}
\multiput(990.00,913.34)(18.301,-4.000){2}{\rule{3.300pt}{0.800pt}}
\put(1022,907.84){\rule{7.709pt}{0.800pt}}
\multiput(1022.00,909.34)(16.000,-3.000){2}{\rule{3.854pt}{0.800pt}}
\multiput(1054.00,906.06)(4.958,-0.560){3}{\rule{5.320pt}{0.135pt}}
\multiput(1054.00,906.34)(20.958,-5.000){2}{\rule{2.660pt}{0.800pt}}
\put(1086,900.34){\rule{7.709pt}{0.800pt}}
\multiput(1086.00,901.34)(16.000,-2.000){2}{\rule{3.854pt}{0.800pt}}
\put(1118,897.34){\rule{6.600pt}{0.800pt}}
\multiput(1118.00,899.34)(18.301,-4.000){2}{\rule{3.300pt}{0.800pt}}
\put(1150,893.34){\rule{6.600pt}{0.800pt}}
\multiput(1150.00,895.34)(18.301,-4.000){2}{\rule{3.300pt}{0.800pt}}
\put(1182,889.84){\rule{7.709pt}{0.800pt}}
\multiput(1182.00,891.34)(16.000,-3.000){2}{\rule{3.854pt}{0.800pt}}
\multiput(1214.00,888.06)(5.126,-0.560){3}{\rule{5.480pt}{0.135pt}}
\multiput(1214.00,888.34)(21.626,-5.000){2}{\rule{2.740pt}{0.800pt}}
\put(1247,882.34){\rule{7.709pt}{0.800pt}}
\multiput(1247.00,883.34)(16.000,-2.000){2}{\rule{3.854pt}{0.800pt}}
\put(1279,879.84){\rule{7.709pt}{0.800pt}}
\multiput(1279.00,881.34)(16.000,-3.000){2}{\rule{3.854pt}{0.800pt}}
\put(1311,876.84){\rule{7.709pt}{0.800pt}}
\multiput(1311.00,878.34)(16.000,-3.000){2}{\rule{3.854pt}{0.800pt}}
\put(1343,873.34){\rule{6.600pt}{0.800pt}}
\multiput(1343.00,875.34)(18.301,-4.000){2}{\rule{3.300pt}{0.800pt}}
\put(1375,869.34){\rule{6.600pt}{0.800pt}}
\multiput(1375.00,871.34)(18.301,-4.000){2}{\rule{3.300pt}{0.800pt}}
\put(1407,866.34){\rule{7.709pt}{0.800pt}}
\multiput(1407.00,867.34)(16.000,-2.000){2}{\rule{3.854pt}{0.800pt}}
\put(380.0,987.0){\rule[-0.400pt]{7.709pt}{0.800pt}}
\sbox{\plotpoint}{\rule[-0.500pt]{1.000pt}{1.000pt}}%
\put(1279,997){\makebox(0,0)[r]{${\footnotesize\sin^22\theta_{e3}}$}}
\multiput(1299,997)(20.756,0.000){5}{\usebox{\plotpoint}}
\put(1399,997){\usebox{\plotpoint}}
\put(220,361){\usebox{\plotpoint}}
\multiput(220,361)(19.229,-7.812){2}{\usebox{\plotpoint}}
\multiput(252,348)(19.811,-6.191){2}{\usebox{\plotpoint}}
\put(298.51,334.37){\usebox{\plotpoint}}
\multiput(316,330)(19.980,-5.619){2}{\usebox{\plotpoint}}
\multiput(348,321)(20.276,-4.435){2}{\usebox{\plotpoint}}
\put(399.29,309.78){\usebox{\plotpoint}}
\multiput(412,307)(20.304,-4.307){2}{\usebox{\plotpoint}}
\put(460.26,297.14){\usebox{\plotpoint}}
\multiput(477,294)(20.400,-3.825){2}{\usebox{\plotpoint}}
\put(521.46,285.66){\usebox{\plotpoint}}
\multiput(541,282)(20.400,-3.825){2}{\usebox{\plotpoint}}
\multiput(573,276)(20.400,-3.825){2}{\usebox{\plotpoint}}
\put(623.64,267.67){\usebox{\plotpoint}}
\multiput(637,266)(20.507,-3.204){2}{\usebox{\plotpoint}}
\put(685.28,258.96){\usebox{\plotpoint}}
\multiput(701,257)(20.276,-4.435){2}{\usebox{\plotpoint}}
\put(746.61,248.72){\usebox{\plotpoint}}
\multiput(765,247)(20.276,-4.435){2}{\usebox{\plotpoint}}
\multiput(797,240)(20.605,-2.498){2}{\usebox{\plotpoint}}
\put(849.16,233.60){\usebox{\plotpoint}}
\multiput(862,232)(20.595,-2.574){2}{\usebox{\plotpoint}}
\put(910.95,225.88){\usebox{\plotpoint}}
\multiput(926,224)(20.400,-3.825){2}{\usebox{\plotpoint}}
\put(972.48,216.64){\usebox{\plotpoint}}
\multiput(990,215)(20.595,-2.574){2}{\usebox{\plotpoint}}
\put(1034.36,209.84){\usebox{\plotpoint}}
\multiput(1054,208)(20.507,-3.204){2}{\usebox{\plotpoint}}
\multiput(1086,203)(20.595,-2.574){2}{\usebox{\plotpoint}}
\put(1137.33,197.19){\usebox{\plotpoint}}
\multiput(1150,196)(20.665,-1.937){2}{\usebox{\plotpoint}}
\put(1199.33,191.38){\usebox{\plotpoint}}
\multiput(1214,190)(20.521,-3.109){2}{\usebox{\plotpoint}}
\put(1261.09,183.68){\usebox{\plotpoint}}
\multiput(1279,182)(20.665,-1.937){2}{\usebox{\plotpoint}}
\put(1323.08,177.87){\usebox{\plotpoint}}
\multiput(1343,176)(20.595,-2.574){2}{\usebox{\plotpoint}}
\multiput(1375,172)(20.665,-1.937){2}{\usebox{\plotpoint}}
\put(1426.30,167.19){\usebox{\plotpoint}}
\put(1439,166){\usebox{\plotpoint}}
\end{picture}
\caption{The numerical results for the ratio of the
    solar neutrino mass squared difference to that
    for the atmospheric neutrino mass squared difference.
    Also shown are the
    squared sine of the double of the solar neutrino
    mixing angle, the atmospheric neutrino
    mixing angle and the mixing angle $\theta_{e3}$.}
  \end{center}
\end{figure}
\indent

In this subsection we will discuss the numerical
calculation. The elements of the mass matrices are
determined, up to factors of order one, to be a product of
several Higgs VEVs measured in units of the fundamental scale
$M_{\sss\rm Planck}$---the Planck scale.

We imagine that the mass matrix elements,~$\eg$~for the
right-handed neutrino masses or for the mass matrix $M_{\nu}^D$,
are given by chain diagrams \cite{fn} consisting of a backbone of
fermion propagators for fermions with fundamental masses,
with side ribs (branches) symbolising a Yukawa coupling to
one of the Higgs field VEVs.

We know neither the Yukawa couplings nor the precise
masses of the fundamental mass fermions. However it is a basic
assumption of the naturalness of our model that these
couplings are of order unity and that, also, the fundamental
masses deviate from the
Planck mass by factors of order unity. In the numerical
evaluation of the consequences of the model, we explicitly take
into account these uncertain factors of order unity by
providing each matrix element with an explicit random number
$\lambda_{ij}$---with a distribution so that its average
$\sVEV{\log {\lambda_{ij}}}\approx 0$ and its spread is
a factor of two.

Then the calculation is performed with these numbers time
after time, with different random number $\lambda_{ij}$-values,
and the results averaged in logarithms. A crude realisation of the
distribution of these $\lambda_{ij}$ could be a flat
distribution between $-2$ and $+2$, then provided also
with a random phase (with flat distribution).

Another ``detail'' is the use of a factor
$\sqrt{{\#} {\rm diagrams}}$ multiplying
the matrix elements, to take into account that, due to the
possibility of permuting the Higgs field attachments in
the chain-diagram, the number of different diagrams is
roughly proportional to the number of such permutations
\mbox{${\#}{\rm diagrams}$}. This is the correction introduced
and studied for the charged fermion mass matrices by D.~Smith and two
of us \cite{fnsnew}. In the philosophy of
each diagram coming with a random order unity factor, the sum of
\mbox{$\#{\rm diagrams}$} different diagrams gets of the order
$\sqrt{{\#} {\rm diagrams}}$ bigger than a
single diagram of that sort\footnote{We have counted these permutations
ignoring the field $S$. If we allowed both $S$ and
$S^{\dagger}$ in the same diagram, the $\sqrt{\#{\rm diagrams}}$
factor could give arbitrarily large numbers if $<S>=1$.
Actually, in the fits with these factorial factors,
even the $S$ and $S^{\dagger}$ are somewhat suppressed and
so the wild blow-up does not occur in practice. Also we
may argue that sub-chains, with equally many $S$
and $S^{\dagger}$ insertions, can be put in all graphs and
their effect is just to renormalize the parameters of the model.}.
It turns out that
these factors are especially important for some elements
involving the electron-neutrino in the matrix
$M_{\nu}^D$, which are suppressed by several factors,
as then many permutations can be made.

Yet another detail is that the symmetric mass matrices---occurring
for the Majorana neutrinos---give rise to the same
off-diagonal term twice in the right-handed neutrino matrix in
the effective Lagrangian. So we must multiply off-diagonal
elements with a factor $1/2$. But in the $M_{\nu}^D$-matrix
columns and rows are related to completely
different Weyl fields and, of course, a similar factor $1/2$ should not
be introduced.

Concerning the $\sqrt{\#{\rm diagrams}}$ factor for the
diagonal mass matrix terms in the symmetric matrices,~$\eg$
{}~$M_R$, we shall remember that (contrary
to what we shall do in non-symmetric matrices such as
$M_{\nu}^D$ and the charged lepton ones) we must count
diagrams with the Higgs fields attachment assigned in
opposite order as only one diagram. The backbone in the
diagram has no orientation and we shall count diagrams
obtained from each other by inverting the sequence of
the attached Higgs fields as only ONE diagram. Thus the
diagonal elements will tend to have only half as many diagrams.

We give in Figure $1$ results, obtained as the average of
$50,000$ random combinations of order unity factors, as a function
of the small VEV $\sVEV{\chi}$ of the new Higgs field $\chi$.
In order to get an atmospheric mixing angle of the order
of unity, the range for $\sVEV{\chi}$ around the ``old''
Anti-GUT VEV $\sVEV{T}\approx0.07$ is suggested; so only
this range is presented.
Finally we present here the orders of magnitude of the right-handed
neutrino masses and the mixing angle $\theta_{e3}$
(see Figure $1$):
\begin{eqnarray}
  \label{eq:right-handedmixinge3}
M_{R_{\nu_{\sss 1}}} &\approx& 10^{11} ~~{\rm\GeV}\nn,\\
M_{R_{\nu_{\sss 2}}} &\approx& 10^{13} ~~{\rm\GeV}\nn, \\
M_{R_{\nu_{\sss 3}}} &\approx& 10^{13}~~{\rm\GeV} \nn,\\
\sin^22\theta_{e3}&\approx& 10^{-4}\nn.
\end{eqnarray}

\section{The problem of scales}

When we, as here, just postulate the see-saw scale to have
the value needed by introducing an appropriate Higgs field,
$\phi_{B-L}$ in our model, and fit its vacuum expectation
value, here to $10^{13}$ $\GeV$, we have just another scale problem.

This new scale problem is of an
exactly analogous character to that of the famous Hierarchy Problem
of why the weak energy scale is so low compared to, say, the Planck
scale (or the unified scale in GUT theories).
So, one could say, we have got one more hierarchy-problem!
It has been popular to identify the see-saw scale with the GUT-scale
but, to do so, one has to make use of the big mass ratios
occurring between the right-handed neutrinos. When it
IS fitted, it is always the heaviest of the see-saw neutrinos that gets
a mass at the GUT-scale. This heaviest right-handed neutrino is that one
which does not have any phenomenological consequences for the
(left-handed) neutrino oscillations, and thus is very model dependent.
To stretch the scale of the see-saw neutrinos from the needed scale, around
the masses presented here of $10^{11}$ to $10^{13}~\GeV$, up to the usual
GUT-scale at $10^{15}$ to $10^{17}~\GeV$ is possible but not
strongly suggested. But, phenomenologically, it is not possible to push the
see-saw scale up to the Planck scale. Thus we must introduce
at least one new scale, in addition to the Planck scale,
the weak scale and the strong scale. This may be very important to
think about in connection with solving the hierarchy problem,
because we now have to look
for a solution that can provide yet another scale---the see-saw
scale or, if we drop the see-saw model altogether,
the neutrino oscillation scale directly.

Preferably we should have a mechanism that could put
these different scales at exponential values relative to, say,
the Planck scale.
This would be much in the same way mathematically
as the strong interaction scale is understood to be
exponential due to the renormalization group running of
$\alpha_S$--the QCD-gauge coupling (squared and divided by 4$\pi$).

We are indeed working on such a project (L.V.~Laperashvili~\cite{Larisanew}
is also included in this project), using the
multiple point principle (MPP) introduced in section
\ref{sec:motivation}. Once we have such a principle that
fixes the couplings and masses in a certain way, we are not really
solving the problem of avoiding finetuning, but rather we have
postulated a \underline{finetuner} that finetunes the parameters
for us, namely the postulated MPP-mechanism. That should make
it much easier to ``solve'' the scale problem or rather
problems---now we include the neutrino oscillations---since
what we need is ``only''
that, when the couplings and masses are being adjusted to make
several degenerate minima as the MPP postulates,
then as a consequence of that
requirement they tend to also organize hierarchies of scales.

The argument on which we work goes roughly like this:

The MPP is taken to postulate that the effective
potential $V_{eff}(\phi_{WS})$
shall have (at least) two minima as a function of
$|\phi_{WS}|^2$, with the same value of the effective potential,
for non-zero values of this Standard
Model Higgs field $\phi_{WS}$. This is
an interpretation of the somewhat vague statement that
there shall be ``many''
minima with the same energy density.
This obviously implies that the second derivative
of this effective potential with respect to $|\phi_{WS}|^2$
has two places on the positive $|\phi_{WS}|^2$
axis where it is positive---namely the minima. It then follows
that there is one place in between where the second derivative
is negative---namely at the maximum that must occur between the
two minima, for topological
reasons so to speak.
For our argument here the degeneracy of the minima does not
seem particularly relevant---it is just that the MPP is a statement
about degenerate minima.

Then we go through, loop-order for loop-order, to see if it is
possible, under reasonable assumptions, that the second derivative
$d^2V_{eff}(|\phi_{WS}|^2)/d(|\phi_{WS}|^2)^2$ could have the
alternating sign behaviour, first positive then negative and
then positive again, required for the presence of two minima.
\vspace{0.1cm}
{\em In 0-loop approximation}:
\begin{equation}
\frac{d^2V_{eff}(|\phi_{WS}|^2)}{d(|\phi_{WS}|^2)^2} = \frac{\lambda}{ 2}
=constant
\end{equation}
$\ie$~it cannot have the required sign shifts except if equal to zero.
\vspace{0.1cm}
{\em In 1-loop approximation}:
\begin{equation}
\frac{d^2V_{eff}(|\phi_{WS}|^2)}{d(|\phi_{WS}|^2)^2} \approx
\frac{\lambda_{run}(|\phi_{WS}|^2)}{ 2}+ constant
\end{equation}
In the approximation that we look for $|\phi_{WS}|^2 \gg \mu^2$
(the Higgs mass squared parameter), the running $\lambda$
is proportional to a single $\log|\phi_{WS}|^2$
with a coefficient that gets positive terms from the bosons
in the field theory and negative terms from the fermions.
Ignoring strange solutions with both minima very
close to each other,
we have a single monotonous logarithm that cannot switch back and forth
in sign. So the only solution is, like in the 0-loop case, to put the
coefficient zero and kill to zero the whole 1-loop contribution
to the second derivative.
\vspace{.1cm}
{\em In 2-loop approximation}:
It is possible to achieve an effective potential second derivative
$d^2V_{eff}(|\phi_{WS}|^2)/d(|\phi_{WS}|^2)$ which can switch
sign as needed to have two minima.

In fact we have found~\cite{FN173} that it is completely possible
to have a scenario with coupling constants and masses leading to
two minima.
It is even completely within the experimentally allowed range
of parameters to have the second minimum 1) at the Planck scale,
and 2) degenerate as the MPP requires. We obtain this particular
scenario when the top mass is $173~\GeV$ and the Higgs mass is
$135~\GeV$, namely the smallest Higgs
mass allowed if the pure Standard Model is valid up to the Planck
scale. Since the indirect estimates of the
Higgs mass favour a mass lower than the experimental limit,
they really suggest it should be
as low as possible and thus just our prediction of
$135~\GeV$, provided no new physics comes in below the Planck
scale allowing a lower Higgs mass.

In fact it is a bit surprising that our calculation, even with a rather
crude value for the Planck scale only assuming its order of magnitude
for the second minimum, yields a top-mass of $173~\GeV$ with an accuracy
similar to the experimental accuracy of about $\pm 6~\GeV$.

So it looks that with two loops we can easily get the two minima
required, but with lower loop accuracy it is impossible. However, the
various couplings are experimentally so small that the two-loop terms
contribution is only a
correction and not usually qualitatively important. If now it is
needed that the two-loop-terms must be important,
in order to have the required two minima
(using MPP), then they must be boosted to importance by sufficiently big
logarithms: In a leading log expansion, the two loop terms
can become comparable or beat out 0-loop and 1-loop approximation
terms if the logs are, so to
speak, as big as the couplings are weak. But the log here must be, for
instance, the logarithm of the ratio of the two minima we required.
If that
gets as big as the couplings are weak, it means that, say,
the lowest minimum field value must go down relative to the
other one by the \underline{exponential} of the inverse couplings.

This would be very similar to what is achieved by means of the
renormalisation group
for the strong scale, but should now numerically also work for the
lower minimum that determines the weak scale. That is potentially a
solution of the scale problem!

There is though a little ``technical'' problem with our
``solution'' of the scale problem: At the second minimum it turns
out, say in our scenario with the Higgs mass being $135~\GeV$,
that the running self-coupling $\lambda_{run}$ for the Higgs
field runs very very small.
If it is allowed to have so small a value, the argument of crude order
of magnitude character given above does not quite work. Then the logical
argumentation for the need of the big ratio of scales fails, even when
the requirement (MPP) of two degenerate vacua is postulated.
Interestingly enough there is a better hope of coming through successfully
with several hierarchy problems at once. So the see-saw scale problem
could turn out being helpful in bringing our idea for solving the scale
problem to work.

\section{Looking for top-quark analogues}
If we accepted the idea that we wanted the postulate of several
minima---or even degenerate minima---to explain large scale ratios,
then we need to have the zero- and one-loop approximation
to cancel approximately~\cite{veltmanstechbled}, so that
the two-loop terms can come to be qualitatively significant.
Remember it was only the two-loop terms that could produce
the required minima.
But the one loop contribution was a logarithm multiplied by positive
terms from the bosons and negative ones from the fermion loops.
This is the reason why we need a strongly coupled top-quark, $\ie$~with
a Yukawa coupling of order unity.

With this very speculative requirement, that each time we have a
Higgs field with a small vacuum expectation value it is due to
our postulated MPP, we can then use it as a guide in seeking
the structure of the model that realizes our assumptions.
As already mentioned the top quark does the job of
cancelling the one-loop correction so as to allow two minima
perfectly, as far as we yet know.

Luckily enough there is, in our above sketched model for the
mass of the see-saw neutrinos coming from the field $\phi_{B-L}$,
just a combination of right-handed neutrinos that can ``play the
role of the top-quark in the Standard Model'', because it couples with
of order unity Yukawa coupling. In fact the Higgs field $\phi_{B-L}$
(which has to have a very small VEV compared to the Planck scale,
in fact of order $10^{13}~\GeV$) couples in our model
to the transition between
the third and the first family right-handed neutrinos. Really it
can convert a right-handed electron neutrino to a CP-antiparticle of the
right-handed $\tau$-neutrino. This coupling is therefore
unsuppressed. That is quite non-trivial in our model, since there can
be so many suppressions of effective couplings due to some charges
needing to be broken to generate the vertex. But nevertheless
we have a top-quark-role-player, needed for the working of our
two minima machinery, for the $\phi_{B-L}$ field.

Now we might say: but even the Higgs fields which in our model
give small hierarchies---the fields T, W, $\xi$, $\chi$ having
only VEV's of the order of $1/10$ in Planck units---strictly
speaking also represent hierarchy-like problems.
Why are their expectation values
so small compared to the fundamental (presumably Planck) scale.
Even a small number of the order of $1/10$ needs an
explanation. So they should also have their ``top-quark-role-players''.
In other words they should have unsuppressed couplings
to some fermion, which is mass protected but gets a mass from
just the Higgs field in question.
Then that fermion could help cancel the one-loop contribution,
so as to make the two-loop one become important.

Such a rule of requiring a ``top-quark-role-player'', for each
Higgs field with very small expectation value relative to the
presumed fundamental scale, will require the existence of some chiral
fermions that then can get mass from the Higgs field in question.
But chiral fermion fields, in turn, will give rise to
anomalies that must be cancelled. So we obtain a series of
consistency requirements that can be useful in building
a model from the bottom up.

We start the construction by seeking Higgs fields to explain the
quark-lepton mass and mixing angle suppressions.
Then, in turn, we look for ``top-quark-role-players'' for these
Higgs fields and, next, for further Weyl-fermions to cancel the anomalies
coming from the ``top-quark-role-players''. The idea then is that the many
consistency requirements shall be used to find a
scheme that works self-consistently.
Then all the small-VEV-Higgs-fields will have their degenerate minima
and fermions, which they give mass to and which play the top quark role.
Furthermore the gauge and mixed anomalies of these fermions will cancel.

One might feel that the story of the ``top-quark-role-players''
has a little too many speculative assumptions to be trustable.
It could then be that there is a slightly different reason
for the same sort of fermion fields to exist---a reason that could
actually be a reformulation of the same model---each Higgs
boson could be imagined and required to be a bound
state of a pair of chiral fermions. That would lead to about the same
requirements as the ``top-quark-player'' rule:
For each low mass (or low VEV) Higgs/scalar field, there must be
some constituents able to form it w.r.t. quantum numbers.
They should typically be fermions and need to be mass protected to
the scale of the mass of the boson to be constructed. The coupling has
to be strong for the constituents to the bound state.

The requirements for a boson being a bound state of a couple of
chiral fermions become very similar indeed to the requirements that
the boson have them as its ``top-quark-role players''.

\section{An extra family of Weyl fermions}

But what fermions could play the ``top-quark-role'' for the
scalar boson fields
$W$, $T$, $\xi$ , $\chi$? It would have to be fermions that actually
obtain a mass from the fields for which they do the job of the top quark.
But that means they end up having masses of the order of the Higgs field
scale in question (the VEV). So we need, in our model, Weyl particles
combining themselves to each other, so as to obtain masses at the
scale of our Higgs fields $W$, $T$, $\xi$ ,... which is
about $1/10\rm{th}$ of the Planck scale.
The mass protection for these Weyl particles
must be due to some of the gauge quantum numbers in our model.
However they should not be mass protected by
the Standard Model quantum numbers,
because then they would have much much lower
masses and perhaps should have been seen.
They must not even have the total $B-L$ as a mass
protecting quantum number, because then they would get
masses of the see-saw scale and not just a factor
of 10 under the Planck scale, as we need them.
They can though have these quantum numbers in a vectorial way,~$\ie$~there
could be equally many right and left Weyl particles with a
given combination of the Standard Model quantum numbers and the
total $B-L$.

For instance we could look at the Higgs field $\chi$,
which was introduced for the sake of fitting the neutrino
oscillations. Since it gives rise to a suppression of the
order of a factor 10 to 15, it must have mass and
VEV at about a factor 10 to 15 under the Planck scale (taken as the
fundamental one). Its quantum numbers in our model are quite simple:
It has no weak hypercharges and then, according to our rule for the
non-abelian quantum numbers, also no non-abelian couplings.
It only has the family-$(B-L)$-quantum numbers:
$(B-L)_2 =-1$ and $(B-L)_3=1$, the rest being zero.

In order for a pair of Weyl fermions to couple to the $\chi$ field
in an unsuppressed way, it is needed that their
quantum numbers add up to those
of the $\chi$ field modulo certain quantum number combinations.
These quantum number combinations correspond to
the ones sitting on the Higgs fields, like the $S$ field in our model,
having expectation values of the order of the fundamental scale.
Otherwise there would be a need for some Higgs field VEV to provide the
lacking quantum number/charge in the coupling and
it would be suppressed. In that case the field would couple too
weakly and the pair of Weyl fermions could not play its top-quark-role.

If we seek to keep the quantum numbers from being too large, the simplest
would be to have one fermion with all quantum numbers zero except for
say $(B-L)_2=-1$, and another fermion with all zero except for
$(B-L)_3=1$. But we can also have that these two Weyl fermions
in addition carry other quantum numbers, but such
that one of these two particles carry just the
opposite further charges to those of the other one.

In fact we are forced to assume that the two particles
that should play the top-quark-role for $\chi$ have
other charges. Otherwise they  would be mass protected
by the total $(B-L)$-charge and get masses
at the see-saw scale, rather than the required
one-tenth of the Planck scale. Indeed they must
each have a ($B-L$)-charge of yet another family so as to get
the total $(B-L)$ be zero. The highly suggested quantum
numbers for these Weyl particles, counted as
left-handed, are thus
\begin{equation}
(0,0,0;-1,0,1)\;\;{\rm and}\;\;(0,0,0;1,-1,0).
\label{start}
\end{equation}
Now is it possible to build such a pair of Weyl particles into a scheme
with mass protected particles cancelling the anomalies and not carrying the
charges that would bring them to the low mass level, where they
would be directly or indirectly seen? That is to say,
we should not mass protect the particles in this scheme
with Standard Model charges. This is most easily done by letting them
all have the total weak hypercharge
$y/2 = y_1/2 + y_2/2 + y_3/2 =0$. If we also want them not
to show up at the see-saw scale, we should not let them
be mass protected by the total
$(B-L) = ( B-L)_1 + (B-L)_2 + (B-L)_3$. Again the most easy
way of ensuring this would be to let
them not have the total $B-L$ at all, $\ie$~to have $B-L=0$ for them.

It is in fact possible to find such a set of Weyl-particles, cancelling
anomalies and containing the two particles suggested. This relatively
elegant scheme of particles contains 15 Weyl particles---just
the same number as the particles in a Standard Model family
without the right-handed neutrino. The quantum numbers of this set
of 15 Weyl particles are listed in Table \ref{fourthfamiliy}.
\begin{table}
\caption{Quantum numbers of added set of 15 Weyl particles.}
\label{fourthfamiliy}
\begin{tabular}{|c|c|c|c|c|c|c|}
\hline\hline
$y_1/2$ & $y_1/2$ & $y_3/2$&$(B-L)_1$&$(B-L)_2$ & $(B-L)_3$&\\
\hline
0&1&-1&0&0&0&{\bf (-1;0)}\\
-1&0&1&0&0&0&{\bf (-2;0)}\\
1&-1&0&0&0&0&{\bf (-3;0)}\\
\hline
0&0&0&0&1&-1&{\bf (0;-1)}\\
0&1&-1&0&1&-1&{\bf (-1;-1)}\\
-1&0&1&0&-1&1&{\bf (-2;1)}\\
1&-1&0&0&-1&1&{\bf (-3;1)}\\
\hline
0&0&0&-1&0&1&{\bf (0;-2)}\\
0&-1&1&-1&0&1&{\bf (1;-2)}\\
1&0&-1&1&0&-1&{\bf (2;2)}\\
-1&1&0&1&0&-1&{\bf (3;2)}\\
\hline
0&0&0&1&-1&0&{\bf (0;-3)}\\
0&-1&1&1&-1&0&{\bf (1;-3)}\\
1&0&-1&-1&1&0&{\bf (2;3)}\\
-1&1&0&-1&1&0&{\bf (3;3)}\\
\hline
\end{tabular}
\end{table}

It can easily be checked that the here proposed particles
have anomalies that cancel among themselves. The last column
in the table contains a shorthand notation for the quantum
numbers, in the way that the first signed number
from 0 to 3 enumerates some simple combinations
of family weak hypercharges, so that:
\begin{eqnarray}
{\bf 0} & = & (0,0,0)\\
{\bf 1} & = & (0,-1,1)\\
{\bf -1} & = & (0,1,-1)\\
{\bf 2}& = & (1,0-1)\\
{\bf -2}& =& (-1,0,1)\\
{\bf 3}& = & (-1,1,0)\\
{\bf -3}& =& ( 1,-1,0)
\end{eqnarray}
The same shorthand notation is used for the ordered sets of the three
family $(B-L)_i$'s. So that~$\eg$~${\bf(-2;1)}$ means
$(y_1/2,y_2/2,y_3/2;(B-L)_1,(B-L)_2, (B-L)_3)$ $=(-1,0,1;0,-1,1)$.

This shorthand notation can be used to recognise a certain amount
of regularity in the found scheme and can be used to quickly see
the cancellation of many of the anomalies.
You should bear in mind that we have so many anomaly constraints
that it is a bit of an art to find any chiral scheme, with
anomaly cancellations for all
the many choices of three charges with potential triangle anomalies.
In fact the scheme is written - moderately - nicely in a
4 by 4 array, as given in Table \ref{array}.

\begin{table}
\caption{The fifteen Weyl particle quantum numbers as an array.}
\begin{tabular}{|c|c|c|c|}\hline
 - & ${\bf (-1;0)}$&${\bf (-2;0)}$&${\bf (-3;0)}$\\
${\bf (0;-1)}$ & ${\bf (-1;-1)}$&${\bf (-2;+1)}$&${\bf (-3;+1)}$\\
${\bf (0;-2)}$ & ${\bf (+1;-2)}$&${\bf (+2;+2)}$&${\bf (+3;+2)}$\\
${\bf (0;-3)}$ & ${\bf (+1;-3)}$&${\bf (+2;+3)}$&${\bf (+3;+3)}$\\
\hline
\end{tabular}
\label{array}
\end{table}

The regularity in this Table is already a little bit complicated:
There are fifteen Weyl particle quantum numbers listed---one
for each of the sixteen combinations, except for the totally sterile
combination ${\bf (0,0)}$ with no charges at all.
These are obtained by combining one of the following
four combinations of a number for the
hypercharges and a sign related to the $(B-L)_i$'s:

\begin{equation}
{\bf [0;-]},\;\;{\bf [1;-]},\;\;{\bf [2;+]},\;\; {\bf [3;+]}
\end{equation}
with one from the following series of a sign related to the
family weak hypercharges and a number for the family $(B-L)_i$'s:

\begin{equation}
{\bf [-;0]},\;\;{\bf [-;1]},\;\; {\bf [+;2]},\;\;{\bf [+;3]}
\end{equation}

The idea here is that, for example, when we combine
${\bf [2;+]}$ from the first series with ${\bf [-;1]}$
from the second series, we construct the
shorthand symbol ${\bf (-2;+1)}$. This then, in turn,
is translated into the combination of charges
supposed to be sitting on that left-handed
Weyl particle with quantum numbers ${\bf (-2;+1)}
=(y_1/2,y_2/2,y_3/2;(B-L)_1,$ $(B-L)_2, (B-L)_3)
=(-1,0,1;0,-1,1)$.

If we now, for example, will check that there are
no anomalies corresponding to a triangle diagram,
with the three external gauge boson couplings being
the ones coupling to the weak hypercharges for family 1 and family 2
and the $(B-L)_2$, it means that we want to check the no anomaly condition
\begin{eqnarray}
& & \sum_{\rm{the~15~Weyl~particles}}
\frac{y_1}{2}\cdot\frac{y_2}{2}\cdot(B-L)_2 \\
&=& \sum_{\sss{\{ {\bf (-3;+1)},{\bf (+3;+3)}\}}}
\frac{y_1}{2}\cdot\frac{y_2}{2}\cdot(B-L)_2 \\
&=& -1 \sum_{ \{ {\bf +1},{\bf +3}\}}(B-L)_2 =0
\end{eqnarray}
It is, namely, easily seen that we can only get contributions
to the product $\frac{y_1}{2}\frac{y_2}{2}$ from the weak
hypercharge shorthand symbols
that are neither having the numbers ${\bf \pm 1}$ nor ${\bf \pm 2}$.
Further to have a $(B-L)_2 $ non-zero factor,
we need to avoid the $(B-L)$-related
shorthand symbol being ${\bf \pm 2}$. In our formulation
this cancellation takes place because the sign
of $y_1y_2/4$ is the same for all the
${\bf \pm 3} = \pm (-1,1,0)$ weak hypercharge combinations,
which are those to which the $y_1y_2/4$
can give a contribution.

In the fifteen Weyl-particle system just presented,
one also finds candidates
for the ``top-quark-role-player'' for a field with
the quantum numbers of the field combination $\xi S^\dagger$,
which has the simple quantum numbers $(0,0,0;1,-1,0)$.

In our fit the Higgs field $S$ has a vacuum expectation value of order
unity. So, in first approximation, there would be essentially no
phenomenological consequences if we replaced the model by a
related one, in which the field $\xi$ had got
the quantum numbers of this combination $\xi S^\dagger$.
Then, when there is use for a Higgs field with the $\xi$ quantum
numbers, one would make use of an extra $S$-field. Since that
does not cost any extra suppression, that would make no great
difference and it could be an equally good model. But it could have
the advantage that we could now find, among the particles which we are
tempted to postulate to exist, some ``top-quark-role-players'' for
the slightly modified $\xi$-field.

But now we must admit that further manipulation of the model
into a related one seems to be needed, if there should be
any chance of getting the suggested requirements of
``top-quark-role-players'', no anomaly fermions and all
that to work. However, that some rudiments of the requirements
come about, by means of almost phenomenologically invented fields,
is a good sign. It suggests that
further investigation in that direction could yield a model that hangs
together and satisfies the requirements.

Let us also remark that this system of fifteen Weyl particles has some
similarity with a usual family of Weyl particles in the Standard Model.
At least it has the similarity that it consists of just 15 particles,
just the number in a usual family. In this way there would be some
sense in talking about such a set of particles as a
``crossing fourth family''. However it is, of course, really quite
misleading, in as far as this ``crossing family'' couples to
completely different gauge fields compared to what a proper fourth
family should do.

\section{On why three families}

It should be noted that the above proposed, and very speculatively
phenomenologically called for, system of $15$ particles, with masses of the
order of a factor ten or so under the Planck mass scale, has some
connection to the question of why we have just three families:

Indeed the construction of a system with just the number of particles
corresponding to a single family, in the way we did it above,
is specific to the case of
just three families! We might think of the abelian
gauge groups, that were used in the construction of the
suggested system of $15$ particles, as
analogous to the Cartan algebra of the Standard Model group. If we
do that, we see that it is natural that we made use of just four
linearly independent abelian charges. Well, at first it looks like we
used $6$ charge-species, namely $3$ weak hypercharges
and 3 family $(B-L)_i$'s.
However, we had to avoid the risk that this system of $15$ particles
gets mass protection by the Standard Model charges or the total
($B-L$)-charge. So we had to let the charge assignments for these $15$
particles
obey the rules that the Standard Model quantum numbers, as well as the
total $(B-L)$-charge, be zero. That imposes two constraints---of
zero $y/2$ and $(B-L)$---upon the charges and there are thus only
$6-2=4$ independent charges.

Now note that, in order that the above construction of the
``crossing fourth family'' should work, we got the condition that the
number of families multiplied by two---one for the
family weak hypercharge $y_i/2$ and
one for the $(B-L)_i$---shall be just two more than the rank of the
Standard Model group.  This constraint can only be satisfied for
the case of just $3$ Standard Model families!

Once we take it as a good idea that there be such a ``crossing fourth
family'', we may again argue for the three families by a slightly different
method: In the following section we shall shortly put forward an idea
of thinking of some especially ``nice'' linear combinations of the
Cartan algebra generators---the ``Han-Nambu-like charges''. We thereby
get an argument for the number
of Weyl-particles in a Standard Model family being just a power of two
minus unity---or simply a power of two
if we include a right-handed neutrino.

Once we observe that there are just $2^4-1 = 15$ particles in a family,
we may ask ourselves whether, by analogy, it would not be expected that
also the number of families and/or the total number of
Weyl particles under the Planck scale should be of the form $2^n-1$.
In this way we are led to propose a picture
of the Weyl fermion charge assignments
for all the particles that are mass protected by some charge in the system:

In addition to the three known families there is a bunch of Weyl-particles,
with the scheme of charge assignments given in Tables $4$ and $5$
above, which are equal in number to those of a Standard Model
family without its right-handed neutrino, namely $15$.
Together this makes $3  (15 +1) + 15 = 63 = 2^6-1$
particles. In other words:
The three known families, each associated to one speculated see-saw
neutrino ( = right-handed neutrino) make up 3(15+1) = 48 Weyl particles.
Adding to them the fantasy extra family -- that is not at all a real
Standard Model family -- suggested by the ``top-quark-role-player''
arguments consisting of 15 particles,
we reach a total of 48+15 = 63. This total of 63 particles can be
considered remarkable by being a power of 2 minus one, just in analogy to
the number of Weyl-particles in a Standard Model family.

But why should these $ 2^n-1$ systems of particles be so important and
likely to occur in nature? We have -- of course --
no really convincing argument for the moment, but
we present below a weak motivation in what is
really an anomaly counting modulo $2$.

\section{Han-Nambu-like charges}

In this section we should like to propose an idea for how one
might make more specific the statement that the
representations of the Standard Model gauge group realized in nature are
{\em very small representations}.

If you look at the non-abelian representations in the Standard Model
you find, except for the gauge fields themselves, {\em nothing but
the trivial representation and the lowest
dimensional representations after the trivial one}.
This can be said to mean that the representations chosen by Nature
are really remarkably {\em small}, as far as the non-abelian groups
$SU(2)$ and $SU(3)$ are concerned!

Can we say the same thing in some sense when we think of abelian group(s)?
For abelian groups you cannot so easily use the dimension of the
representation, since abelian groups always have one-dimensional
representations and thus dimension makes no distinction; instead the
natural suggestion would be to use the ratio of the charges
of the representation realized compared to the quantum of charge---the
Millikan-like charge value of which all charges must be an integer
multiple. One might now think of looking at the unique abelian {\em invariant}
subgroup in the Standard Model gauge group, the weak hypercharge;
but taken literally that idea does {\em not} work well, if
you hope to show that the Standard Model charges are remarkably small.
The problem is that the quantum for
half the weak hypercharge $y/2$ is 1/6, while the
right-handed charged leptons have the value $y/2 = -1$. So it is
$6$ times as large, but $6$ is not the smallest integer after zero!

However there may be no good reason for looking only at the invariant
abelian subgroups. So we suggest taking into account also the
non-invariant abelian subgroups. But there are a lot of abelian subgroups
of the Standard Model group; really the Standard Model group has infinitely
many abelian subgroups of course.
Think for example of
two non-invariant abelian subgroups of the Standard Model
group, such as the one generated by the electric charge and the
one generated by the Gell-Mann $\lambda_8$ or say $\sqrt{3}\lambda_8$.
The electric charge in the Standard Model really has a quantum equal to
$1/3$ in units of the Millikan quantum, because the quarks have
non-integer charges, while  $\sqrt{3}\lambda_8={\rm diag} (1, 1, -2)$ has
a quantum of charge equal to $1$. If we, however, now
make a linear combination with complicated rational coefficients,
we easily get a combined charge or generator in the Standard Model Lie algebra
which can have a very small quantum of charge. So it is easy, by forming
linear combinations of Lie algebra generators (essentially
the same as charges),
to get such generators that have very small quanta of charge
and for which the actual particles then
turn out to have a huge number of quanta.
Thus, w.r.t. such abelian charges, the Standard Model
representations will be``very big ''~$\ie$~have lots of quanta for
the realized representations ($\approx$ particles). In formulating a statement
about whether the Standard Model has small or big abelian representations,
it is therefore necessary to keep in mind that the different non-invariant
charges we can think of have very different charge quanta sizes, in
general, and thus
are not at all represented on equally ``small'' representations.

These complicated linear combinations (let alone linear combinations with
irrational coefficients) are not so attractive to study or use for the
definition of the representations being small or large. So we prefer
to define the question of whether a model---say
the Standard Model---has small or large representations
in terms of whether it is possible to define many, or only a few, charges
which are realized with very low numbers of charge quanta.
The trivial and smallest non-trivial charges you can hope for
are $0$ or $1$ or $-1$ measured in charge quanta.  So we
should say that the abelian representations
being small should, by definition, mean
that they have relatively many charges ($\ie$~generators in the Lie algebra)
with the property of having only the values $0$, $1$ or $-1$ realized.

So the question then is: will a reasonable ``counting'' of the
charges (Lie algebra generators), that have this property of only
having the three possibilities $0$, $1$ and $-1$ for
the number of quanta on a particle, lead to the Standard Model having
especially many such charges?
The answer we want to suggest is that it is indeed so that the Standard
Model has a relatively big family of such charges. Then one should be
able to say that w.r.t.~the abelian charges (and not only w.r.t.
the non-abelian ones) the Standard Model is a model with ``small''
representations!

Let us, for simplicity, consider the Cartan algebra of the Standard Model
gauge group
and look for how many
Lie algebra generators (or charges) we can find,
inside this Cartan algebra,
which have the property of having only eigenvalues $1,0,-1$.

Actually we can present a family of ten such generators, several
of which are already known in the literature as variations of the
Han-Nambu charges \cite{HanNambu}. Remember that the Han-Nambu
charges were proposals for what the electric charge could be in a model
for quarks etc., which lost support because of its disagreement with
the results from deep inelastic scattering experiments. This model
has the property that even the quarks have integer Millikan
quanta of charges of this Han-Nambu type (which was suggested as
a candidate for electric charge, but here is just considered a certain
formal charge we can construct if we like). In fact, the idea is that
one adds to the by now generally accepted form of the electric charge
in the Standard Model, $Q=y/2 + t_3$, a colour dependent term
$\lambda_8/\sqrt{3}$  which has eigenvalues $1/3$, $1/3$ and $-2/3$. If
the quark which has the $-2/3$ eigenvalue is a red one,
we can say we get the ``red Han-Nambu charge''.

It is clear that, even inside the Cartan algebra chosen by using
diagonal $t_3$ and Gell-Mann's $\lambda_8$ and $\lambda_3$, we can find
Han-Nambu charges of the two other colours--the blue and the yellow.
This already makes up three Han-Nambu charges inside the Cartan algebra.
In addition we can take ``anti-Han-Nambu charges'', which
are obtained by replacing the electric charge $y/2 + t_3$ by
the the charge $y/2 - t_3$, $\ie$ with a minus sign on the
$t_3$-term. This gives us three more charges which,
again, are easily seen to
have no other eigenvalues than $0,1,-1$. Now it is actually easily understood
that if we have two charges, say $Q_A$ and $Q_B$, which have only the three
eigenvalues $-1$, $0$ and $1$, then there is an enhanced chance that the sum or
the difference would again be a charge with this property.
It is, of course, by
no means guaranteed, since we could easily risk that say the sum $Q_A+Q_B$
has also double charges.  However it turns out that we can indeed construct
sums and differences of some of the Han-Nambu and anti-Han-Nambu charges
which again have this $-1,0,1$ property. For instance subtraction of
the Han-Nambu charges from each other leads to $\lambda_3$ type colour
generators, of which we have three combinations in the Cartan algebra
(counting as one a charge and its opposite).
By subtracting analogous
anti-Han-Nambu and Han-Nambu charges, we can also get $2t_3$ which is
$1$, or $-1$ on left-handed particles and $0$ on the right-handed (in the
conventional thinking of particles but ignoring of anti-particles).

This makes up the following $10$ charges in the Cartan algebra with our
$-1,0,1$ property:

\begin{eqnarray}
Q_{\rm HN\; red} &=& y/2 + t_3/2 + \lambda_{8(\rm{red})}/\sqrt{3}\\
Q_{\rm HN\; blue} &=& y/2 + t_3/2 + \lambda_{8(\rm{blue})}/\sqrt{3}\\
Q_{\rm HN\; yellow} &=& y/2 + t_3/2 + \lambda_{8(\rm{yellow})}/\sqrt{3}\\
Q_{\rm \overline{HN}\; red} &=& y/2 - t_3/2 +
\lambda_{8(\rm{red})}/\sqrt{3}\\
Q_{\rm \overline{HN}\; blue} &=& y/2 - t_3/2 +
\lambda_{8(\rm{blue})}/\sqrt{3}\\
Q_{\rm \overline{HN}\; yellow}&=& y/2 - t_3/2 +
\lambda_{8(\rm{yellow})}/\sqrt{3}\\
\lambda_{3\,\rm{red,~blue}}\\
\lambda_{3\,\rm{blue,~yellow}}\\
\lambda_{3\,\rm{yellow,~red}}\\
2t_3
\end{eqnarray}

If the sum of two of these $10$ charges happens to be one of the
other ones, the difference will not be one
but will have double charge eigenvalues.
They also form an algebra, together with $5$ generators
that have eigenvalues $\pm 2$ though, which
is to be considered a modulo $2$ algebra. This is to avoid having
to distinguish between sums and differences, in order to really
have the set of these charges closed under addition (because sometimes
it is the sum sometimes the difference that belongs again to the set).

\section{Modulo $2$ considerations}

In the light of the fact that the algebra of the Han-Nambu-like charges
tends to be a modulo $2$ algebra, it is natural to consider the
set of Han-Nambu-like charges as a $Z_2$-vector space. We may
then first study the anomaly restrictions
for the $Z_2$-vector space, hoping that this study is
simpler but can nevertheless give interesting information.

Thus we shall study here the anomaly restrictions
on a set of general Han-Nambu-like charges having,
to as large an extent as possible, only the charges
$-1,0,1$ on the particles. By such a study we hope
to partially derive the Standard Model pattern of particles.
So let us here imagine a series of ``Han-Nambu-like charges'' $Q_{HN\; i}$
$(i = 1,2,3,...,n)$,
having the property that most of the charges take on the values
$Q_{HN\; i}=-1,0,1$ for all the particles in the general model under
consideration.
In $Z_2$ algebra language, we throw away the information of whether the
charge is $1$ or $-1$ but keep the information of whether it is
even or odd.

We now consider the triangle anomaly cancellation conditions
\begin{equation}
\sum_{\rm the \;\; Weyl\;\; particles}Q_{HN\; i}Q_{HN\; j} Q_{HN\; k} =0
\label{namod2}
\end{equation}
and the mixed anomaly condition
\begin{equation}
\sum_{\rm the \;\; Weyl\;\; particles}Q_{HN\; i} =0
\end{equation}
for the model. These conditions can be interpreted as applying to the
true gauge charges on the particles or as only applying to the charges
modulo $2$,
so that the symbols $Q_{HN\; i}$ only take the values ``even'' or ``odd''.
Then of course one shall use the algebra of the field $Z_2$. Considered
this $Z_2$ way the mixed anomaly condition is really superfluous, as it
is a special case of eq.~(\ref{namod2}) with $i=j=k$.

We now successively introduce a few assumptions that are supposed to be
consequences or more precise forms of ``the requirement of
small representations''. Firstly this requirement should
mean that there exist many Han-Nambu-like charges. This in turn
suggests that when we combine two Han-Nambu-like charges, by
addition or subtraction, then we get another Han-Nambu-like
charge (having only eigenvalues -1, 0, +1), in as many cases
as possible. In the cases when the combination is not a
Han-Nambu-like charge, the resulting charge has eigenvalues
$\pm 2$ as well. Next we shall make this requirement of often
finding the sum or difference to be a new Han-Nambu-like charge
more specific as follows: Gauge field theory models with
the ``smallest possible representations'' contain a
group (algebra) of charges closed under $Z_2$-addition, of
which as many as possible are Han-Nambu-like (having
eigenvalues $\pm 1$, $0$ only), while as few as possible
may in addition have eigenvalues $\pm 2$.

{}From the point of view of the $Z_2$-algebra, the Weyl fermions of the
model have their charges in the vector space dual
to the charges $Q_{HN\; i}$:
given a particle $ p$ and a charge we get an inner product
$\left\langle Q_{HN\; i}|p \right\rangle$ =
``the (eigen)value of the charge $Q_{HN\; i}$ for the particle $p$'' = ``the
charge of the particle p of kind $Q_{HN\; i}$''.
Again we may think of this algebra as being counted modulo $2$ if we like
to do so, and now we do just that.

It would be most elegant, and we could also claim that it would
mean the biggest number of charges of this ``Han-Nambu-like'' type,
if we had so many that we used up all modulo $2$ independent charge
assignments.
Let us take this as an excuse for assuming that the mass protection
of the particles we consider---we are after all
interested in observable low energy physics particles which
should be mass protected---is revealed in the modulo $2$ counting.
In order for a mass protected particle to exist with
some combination of the charges $Q_{HN\;i}$ specified ($\ie$~in a
certain vector of the dual vector space of the space of these charges),
it is necessary  to have an odd number of particle types (flavours)
with these charges modulo 2.

Interpreting the no-anomaly condition (\ref{namod2}) as a $Z_2$-charge
condition, the term $Q_{HN\; i}Q_{HN\; j}Q_{HN\; k}$ in the sum is
only odd ($\ie$ non-zero) when all three charges $Q_{HN\; i}$,
$Q_{HN\; j}$ and $Q_{HN\; k}$ are odd for the Weyl particle
considered. Expressed geometrically, this means that one only
gets non-zero contributions from particles which are associated
with the intersection of the three ``displaced hyperplanes''
$\left\{ p \in V_p \mid Q_{HN\; i}(p) = \mbox{odd}\right\}$,
$\left\{ p \in V_p \mid Q_{HN\; j}(p) = \mbox{odd} \right\}$
and $\left\{ p \in V_p \mid Q_{HN\; k}(p) = \mbox{odd} \right\}$
in the $Z_2$-vector space $V_p$ of particles.
If the $Z_2$-vector spaces are of dimension
less than or equal to three, this intersection can be made at just
one point and if all the charges are gauged we can make it any point.
So it is impossible to obtain a cancellation of anomalies between
particles and all the charges must be even, giving no
particles mass-protected modulo 2, if the charge and particle space is
of dimension less than or equal to $3$.

Thus the smallest allowed dimension for the space of
particles, as well as that of the charges $Q_{HN\;i}$, should be four.
In this case one can easily show that, if just one charge combination
is assumed to have an odd number of particles, then all non-trivial charge
combinations (we now only work modulo 2) have an odd number of particles.
Of course a particle with all charges $0$ is of no help in solving the anomaly
conditions and can be left out. This means that we have to fill $2^4-1=15$
possible charge assignments modulo $2$ with an odd number of particles.

The number $15$ is interesting: there are just $15$ Weyl particles in
a Standard Model family, and the $10$ (good with only $-1,0,1$) + $5$
(bad, with double charges) = $15$ Han-Nambu-like charges in the
Standard Model can be seen to be assigned to the $15$ particles in a
family.
These $15$ Weyl particles are distributed over all the 15
non-zero vectors in the
particle-vector-space dual to that of the charges!

In other words, with a few ``smallest representation assumptions'',
the Standard Model structure modulo $2$  for the Han-Nambu-like
charges is obtained as the smallest
dimensional $Z_2$-space that can cancel the
anomalies but still be mass protected!
Crudely speaking this means we can claim
that indeed a Standard Model family
is having the smallest representation, as far as the charges in the Cartan
algebra goes.

We shall now take the above argumentation to suggest more generally a
$Z_2$-vector space structure, for both a set of Han-Nambu-like charges
and the set of Weyl particles. It is then suggested that we should find,
in the true model of Nature, a number of Weyl particles
equal to a power of $2$
perhaps minus $1$. The latter subtraction is expected,
because we can in no way
mass protect a super-sterile particle having all its quantum numbers zero.
So the particles that are totally ``even'' in the $Z_2$-formulation
may be totally sterile and impossible to mass protect. Thus we
would count them as only belonging to the ``garbage at the Planck scale'',
where everything is to be found in our philosophy.

The number of already observed Weyl fermions is, of course,
$45$ corresponding to the three Standard Model families. This
number could be increased to 48 if one took seriously the
indirect evidence for three see-saw scale right-handed neutrinos.
The next power of 2 minus 1, greater than 48, is 63.
So, if the above idea were to be upheld, there would be a need
to postulate the existence of
an extra set of $15$ Weyl particles. This extra set of particles
would have to be so weakly mass protected
that the particles have become so heavy that we do not see them.
There is an extra scale in our Anti-GUT
model a single order of magnitude under the Planck scale. So it is,
of course, not unexpected that there could be Weyl particles
in the Anti-GUT model mass protected only down to this scale---counting
the Planck scale as the {\it a priori} mass scale.
Indeed one could easily imagine constructing a system of $15$ particles,
in analogy with a usual Standard Model family,
just using some other
gauge fields with a more strongly broken/Higgsed gauge symmetry.
If this extra set of particles have masses of the order
of $1/10$ of the Planck mass,
they will stay safely in the fantasy sector for
a long time to come!

Once you work with $Z_2$ vector spaces,
you can imagine that somehow or another the known Standard Model
families fit into a $Z_2$ vector space with four elements, $\ie$ of
dimension $2$. It should then be remarked
that there is a permutation symmetry of the
$Z_2$ vector field structure among the three non-zero elements while the
zero-vector is, of course, special w.r.t.~the $Z_2$-algebra. This
could be interpreted as support for the existence of 3 families.

\section{Conclusion}
\label{sec:conclusion}

\indent
In this article we have made an extension of the Anti-GUT
model to neutrinos, by including see-saw
$\nu_R$ particles at a scale of mass around $10^{12}~\GeV$.
By this extension we
introduced two more parameters, namely the vacuum expectation
values of two additional Higgs fields, $\phi_{\sss B-L}$ and
$\chi$. But one extracts
two mixing angles, $\theta_{\odot}$ and $\theta_{\rm atm}$, and
two mass squared differences, $\Delta m^2_{\odot}$, and
${\Delta m^2_{\rm atm}}$ from the neutrino oscillation data.
So in this sense we have two predictions:
\begin{eqnarray}
  \label{eq:ergebnis}
  \sin^22\theta_{\odot} &\approx& 3 \times 10^{-2}\\
  \frac{\Delta m^2_{\odot}}{\Delta m^2_{\rm atm}} &\approx& 6 \times 10^{-4}
\end{eqnarray}

These results are {\em only order of magnitude} estimates, and we shall
count something like an uncertainty of $50\%$ for mixing angles
and masses. Thus for the square, $\sin^22\theta$, we estimate
an uncertainty of $100\%$~$\ie$~a
factor of $2$ up or down and for the ratio,
${\Delta m^2_{\odot}}/{\Delta m^2_{\rm atm}}$,
an uncertainty of $\sqrt{2}\cdot 100\%$
meaning roughly a factor of $3$ up or down:
\begin{eqnarray}
  \label{eq:ergebnis2}
    \sin^22\theta_{\odot} &=& (3 {\sss{ +3\atop -2 }})\times 10^{-2}\\
    \frac{\Delta m^2_{\odot}}{\Delta m^2_{\rm atm}} &=&
(6 {\sss{ +11\atop -4 }})\times 10^{-4}\nn.
\end{eqnarray}%

These two small numbers both come from the parameter
$\xi$---the VEV in ``fundamental units'' of one of
the $7$ Higgs fields in our model---which has
already been fitted to the charged fermions in earlier
works \cite{mark1}. Its value, eq.~(\ref{eq:vevs}),
is essentially the Cabibbo angle measuring the
strange to up-quark weak transitions ($\xi\simeq0.1$ essentially
giving $\sin\theta_{c}\simeq 0.22$). More precisely we find
from our fit \cite{NT}:
\begin{eqnarray}
  \sin^22\theta_{\odot} &=& 3\,\xi^2\label{eq:ergmitxi1}\\
  \sin\theta_{c} &=& 1.8\,\xi\label{eq:ergmitxi2}\\
  \frac{\Delta m^2_{\odot}}{\Delta m^2_{\rm atm}} &=&
6\,\xi^4\label{eq:ergmitxi3}
\end{eqnarray}
The numerical
factors in front of equations (\ref{eq:ergmitxi1}), (\ref{eq:ergmitxi2})
and (\ref{eq:ergmitxi3}) are the result of
our rather arbitrary averaging over random order unity factors and
the inclusion of diagram counting square root factors
(factorial corrections), as put
forward in reference \cite{fnsnew}. But in principle these
factors are just of order unity.
It is also important
for the success of our model that there has been room to
put in the $\chi$ field, with which we could fix the
atmospheric mixing angle to be of order unity (by taking
$\chi\sim T$), as well as a parameter $\phi_{\sss B-L}$.
The latter parameter is the Higgs field VEV for
breaking the gauged $B-L$ charge and is used
to fit the overall scale of the	observed neutrino masses.

\begin{table}[!!b]
\caption{Number of parameters.}
\label{Table3}
\begin{center}
\begin{tabular}{|c||ccc|c|c|} \hline
& \hspace{2mm}{\small ``Yukawa''}&\vline
&{\small ``Neutrino''}\hspace{2mm} &
{\small {\#} of parameters} & {\small {\#} of predictions}    \\ \hline
{\small Standard Model} & \hspace{2mm} $13$
& \vline & $4$ \hspace{2mm} & $17$  & ---\\ \hline
{\small ``Old'' Anti-GUT}   &  \hspace{2mm}  $4$ & \vline
& ---$\sss{{^\ast}}$ \hspace{2mm} & $4^{\sss{\dagger}}$ & $9$ \\ \hline
{\small ``New'' Anti-GUT}   & & $6$ & & $6$ & $11$ \\ \hline
\end{tabular}
\end{center}
\vspace*{-0.4cm}
\begin{center}
{\small $^\ast$ The ``old''Anti-GUT cannot predict the neutrino oscillation.\\
$^{\dagger}$ Here we have not counted the neutrino oscillation parameters.}
\end{center}
\end{table}

We want to emphasise here that our model---extended Anti-GUT
as well as ``old'' Anti-GUT---is itself a good model in the
sense that all coupling constants are of order of unity,
except for Higgs fields VEVs and thereby also the Higgs masses
giving rise to these VEVs. In the Standard Model the most remarkable
unnatural feature is the tremendously
small value of the Weinberg-Salam Higgs VEV compared to the Planck or
realistic GUT scales. If somehow we have to accept that
there must be a mechanism in nature for making the Weinberg-Salam
Higgs VEV very small, we also should admit that the other Higgs VEVs
could be very small. In our model we manage to interpret
the  second unnatural feature of most Standard Model
Yukawa couplings, namely
that they are very small, to be also due to small Higgs field VEVs.
In this way all small numbers come from VEVs in our model;
the rest is put to unity in Planck units. In this sense it is
``natural'', meaning that it has only one source of small
numbers---the VEVs.
Even the gauge couplings can be interpreted as being of order of unity,
if we follow our MPP assumption, which goes together extremely
well with the present model.

Concerning this question of the Higgs fields often having numerically
small values compared to the fundamental (Planck) scale,
we presented an idea for how these different scales
could come in order of magnitudewise: The proposal was that there is
a principle working in nature, which finetunes the coupling constants so as
to arrange for several degenerate vacuum states.
We argued that that there is a good chance that
exponentially small VEV-scales could come out from this just postulated
finetuning principle (called MPP = multiple point principle).
Such a picture pre-supposes some dynamics, so to speak, in the
coupling constants much like in baby universe theory~\cite{baby}.

So we may say that we did not really solve the hierarchy problem
in the sense of getting the scale out {\em without } finetuning, but
rather proposed a model for finetuning, namely the MPP. We have earlier
claimed some success with such a principle w.r.t. getting finestructure
constants and predicting the top-mass to be $173\pm6~\GeV$.
Also there is a
true \underline{prediction} for the Higgs mass of $135\pm9~\GeV$
(so watch out in the future).

For the working of our scheme to get scales of highly different
orders of magnitude, it is important to have a fermion, like
the top quark in the Standard Model, with a large unsuppressed
coupling to the relevant Higgs field---$\eg$ to finetune the Weinberg-Salam
Higgs field VEV to the electroweak scale rather than the Planck scale.
Assuming that we need a similar unsuppressed fermion for
the other cases of Higgs fields in the Anti-GUT model,
some heavy fermions at ~$\eg$~$1/10$th the Planck scale mass
are very speculatively required to exist.
For this purpose, we proposed a set of $15$ Weyl particles, with
anomaly cancellation and carrying family $(B-L)_i$ charges
and weak hypercharges
only. Since they are not to be mass protected by the Standard Model,
these $15$ very heavy particles must have
zero diagonal quantum numbers,~$\ie$~with zero total $(B-L)$
and zero total $y/2$.

In this connection we had some weak indication that
the number 3 for the number of families had some special significance
in these constructions. We also suggested that the Standard
Model could be considered as having
very small, one could say minimal, representations
w.r.t.~(non-invariant) abelian (sub)groups, as well as
for the non-abelian groups.

At the end we should emphasize
that our Anti-GUT model mass matrices have managed to
fit order of magnitudewise about
$17$ quantities ($11$ observed fermion masses or mass
squared differences, $5$
mixing angles and the CP-violating phase of quarks) with $6$
parameters---the Higgs field VEVs.
As we can see from Table \ref{Table3}, in order to fit the
four quantities measured in the neutrino oscillation data,
we have introduced two more parameters and gained
two predictions (the solar mixing angle and the ratio
of the neutrino oscillation masses).
Also it should be mentioned that we really developed this Anti-GUT model, by
seeking to understand the fermion spectrum in a model that
D.L.~Bennett and I.~Picek used with one of us (H.B.N.) to predict the
values of the fine structure
constants. The postulates of this earlier model were
later replaced by the the above-mentioned MPP.
These predictions of the finestructure constants
actually agree with experiment, with an uncertainty
of $\pm 6$ in the inverse fine structure constants \cite{glasgow}.
When you remember
that this same MPP is promising for some finetuning problems, we can claim
that the combined MPP and (extended) Anti-GUT model provides a fit to
a very large number of the Standard Model parameters!

Note that our model is very successful in describing neutrino
oscillations and their mixing angles, but this model does not
have any good candidate for dark matter; the monopoles could be
such a candidate. We will study this problem in a forthcoming article.

\section*{Acknowledgements}

We would like to thank M.~Gibson and S.~Lola for useful
discussions. Two of us (H.B.N. and Y.T.) wish to thank
W.~Buchm{\"u}ller and T.~Yanagida for an important discussion
of the see-saw mechanism.  S.E.~Rugh and D.L.~Bennett are thanked
for discussions concerning Han-Nambu charges, and L.V.~Laperashvili
especially concerning the problem of scales.
H.B.N. wishes to thank the EU commission for grants
SCI-0430-C (TSTS) and CHRX-CT-94-0621. C.D.F. and H.B.N.
thank the EU commission for grants
INTAS-RFBR-95-0567 and INTAS 93-3316(ext).
C.D.F. also thanks PPARC for its support of this research. Y.T. thanks
the Scandinavia-Japan Sasakawa foundation for grants No.00-22
and  the Theory Division of CERN for the hospitality extended to him
during his visits.

\end{document}